\documentclass[twocolumn]{aastex631}

\usepackage{enumitem}
\usepackage{physics}
\usepackage{multirow}
\usepackage{hhline}
\usepackage{chngcntr}
\usepackage{hyperref}

\hypersetup{pdfnewwindow=true,
     colorlinks=true, 
     linkcolor=blue,
     citecolor=blue,
	 final=true}
	 
\usepackage{etoolbox}
\makeatletter
\patchcmd{\NAT@citex}
  {\@citea\NAT@hyper@{%
     \NAT@nmfmt{\NAT@nm}%
     \hyper@natlinkbreak{\NAT@aysep\NAT@spacechar}{\@citeb\@extra@b@citeb}%
     \NAT@date}}
  {\@citea\NAT@nmfmt{\NAT@nm}%
   \NAT@aysep\NAT@spacechar\NAT@hyper@{\NAT@date}}{}{}
\patchcmd{\NAT@citex}
  {\@citea\NAT@hyper@{%
     \NAT@nmfmt{\NAT@nm}%
     \hyper@natlinkbreak{\NAT@spacechar\NAT@@open\if*#1*\else#1\NAT@spacechar\fi}%
       {\@citeb\@extra@b@citeb}%
     \NAT@date}}
  {\@citea\NAT@nmfmt{\NAT@nm}%
   \NAT@spacechar\NAT@@open\if*#1*\else#1\NAT@spacechar\fi\NAT@hyper@{\NAT@date}}
  {}{}
\makeatother

\newcommand{\JWST}{\textit{JWST}}
\newcommand{\HST}{\textit{HST}}

\newcommand{\COS}{COS-z8M1}
\newcommand{\CEERS}{CEERS-z7M1}

\newcommand{\hii}{H\,\textsc{ii}}
\newcommand{\nii}{[N\,\textsc{ii}]}
\newcommand{\oiii}{[O\,\textsc{iii}]}
\newcommand{\prospector}{\textsc{prospector}}
\newcommand{\bagpipes}{\textsc{bagpipes}}
\newcommand{\cigale}{\textsc{cigale}}
\newcommand{\beagle}{\textsc{beagle}}

\begin{document}

\title{\large Two massive, compact, and dust-obscured candidate \\[0.15em] $\boldsymbol{z\simeq 8}$ galaxies discovered by \textit{JWST}}

\shortauthors{Akins et al.}
\shorttitle{Massive, dust-obscured candidate $z\simeq 8$ galaxies}

\suppressAffiliations
\correspondingauthor{Hollis B. Akins} 
\email{hollis.akins@gmail.com}

\author[0000-0003-3596-8794]{Hollis B. Akins}
\affiliation{Department of Astronomy, The University of Texas at Austin, 2515 Speedway Blvd Stop C1400, Austin, TX 78712, USA}

\author[0000-0002-0930-6466]{Caitlin M. Casey}
\affiliation{Department of Astronomy, The University of Texas at Austin, 2515 Speedway Blvd Stop C1400, Austin, TX 78712, USA}
\affiliation{Cosmic Dawn Center (DAWN), Denmark}

\author[0000-0001-9610-7950]{Natalie Allen}
\affiliation{Cosmic Dawn Center (DAWN), Denmark}
\affiliation{Niels Bohr Institute, University of Copenhagen, Jagtvej 128, DK-2200, Copenhagen N, Denmark}

\author[0000-0002-9921-9218]{Micaela B. Bagley} 
\affiliation{Department of Astronomy, The University of Texas at Austin, 2515 Speedway Blvd Stop C1400, Austin, TX 78712, USA}

\author[0000-0001-5414-5131]{Mark Dickinson} 
\affiliation{NSF’s National Optical-Infrared Astronomy Research Laboratory, 950 N. Cherry Ave., Tucson, AZ 85719, USA}

\author[0000-0001-8519-1130]{Steven L. Finkelstein}  
\affiliation{Department of Astronomy, The University of Texas at Austin, 2515 Speedway Blvd Stop C1400, Austin, TX 78712, USA}

\author[0000-0002-3560-8599]{Maximilien Franco} 
\affiliation{Department of Astronomy, The University of Texas at Austin, 2515 Speedway Blvd Stop C1400, Austin, TX 78712, USA}

\author[0000-0003-0129-2079]{Santosh Harish} 
\affiliation{Laboratory for Multiwavelength Astrophysics, School of Physics and Astronomy, Rochester Institute of Technology, 84 Lomb Memorial Drive, Rochester, NY 14623, USA}

\author[0000-0002-7959-8783]{Pablo Arrabal Haro} 
\affiliation{NSF’s National Optical-Infrared Astronomy Research Laboratory, 950 N. Cherry Ave., Tucson, AZ 85719, USA}

\author[0000-0002-7303-4397]{Olivier Ilbert}
\affiliation{Aix Marseille Université, CNRS, CNES, LAM, Marseille, France}

\author[0000-0001-9187-3605]{Jeyhan S. Kartaltepe} 
\affiliation{Laboratory for Multiwavelength Astrophysics, School of Physics and Astronomy, Rochester Institute of Technology, 84 Lomb Memorial Drive, Rochester, NY 14623, USA}

\author[0000-0002-6610-2048]{Anton M. Koekemoer}
\affiliation{Space Telescope Science Institute, 3700 San Martin Drive, Baltimore, MD 21218, USA}

\author[0000-0001-9773-7479]{Daizhong Liu}  
\affiliation{Max-Planck-Institut für Extraterrestrische Physik (MPE), Giessenbachstr. 1, D-85748 Garching, Germany}

\author[0000-0002-7530-8857]{Arianna S. Long}
\altaffiliation{Hubble Fellow}
\affiliation{Department of Astronomy, The University of Texas at Austin, 2515 Speedway Blvd Stop C1400, Austin, TX 78712, USA}

\author[0000-0002-9489-7765]{Henry Joy McCracken}
\affiliation{Institut d’Astrophysique de Paris, UMR 7095, CNRS, and Sorbonne Université, 98 bis boulevard Arago, F-75014 Paris, France}

\author[0000-0003-2397-0360]{Louise Paquereau}
\affiliation{Institut d’Astrophysique de Paris, UMR 7095, CNRS, and Sorbonne Université, 98 bis boulevard Arago, F-75014 Paris, France}

\author[0000-0001-7503-8482]{Casey Papovich} 
\affiliation{Department of Physics and Astronomy, Texas A\&M University, College Station, TX, 77843-4242 USA}
\affiliation{George P. and Cynthia Woods Mitchell Institute for Fundamental Physics and Astronomy, Texas A\&M University, College Station, TX, 77843-4242 USA}

\author[0000-0003-3382-5941]{Nor Pirzkal}
\affiliation{ESA/AURA Space Telescope Science Institute}

\author[0000-0002-4485-8549]{Jason Rhodes}
\affiliation{Jet Propulsion Laboratory, California Institute of Technology, 4800 Oak Grove Drive, Pasadena, CA 91001, USA}

\author[0000-0002-4271-0364]{Brant E. Robertson}
\affiliation{Department of Astronomy and Astrophysics, University of California, Santa Cruz, 1156 High Street, Santa Cruz, CA 95064, USA}

\author[0000-0002-7087-0701]{Marko Shuntov}
\affiliation{Cosmic Dawn Center (DAWN), Denmark}
\affiliation{Niels Bohr Institute, University of Copenhagen, Jagtvej 128, DK-2200, Copenhagen N, Denmark}

\author[0000-0003-3631-7176]{Sune Toft}
\affiliation{Cosmic Dawn Center (DAWN), Denmark}
\affiliation{Niels Bohr Institute, University of Copenhagen, Jagtvej 128, DK-2200, Copenhagen N, Denmark}

\author[0000-0001-8835-7722]{Guang Yang} 
\affiliation{Kapteyn Astronomical Institute, University of Groningen, P.O. Box 800, 9700 AV Groningen, The Netherlands}
\affiliation{SRON Netherlands Institute for Space Research, Postbus 800, 9700 AV Groningen, The Netherlands}

\author[0000-0001-6813-875X]{Guillermo Barro}
\affiliation{Department of Physics, University of the Pacific, Stockton, CA 90340, USA}
\author[0000-0003-0492-4924]{Laura Bisigello}
\affiliation{INAF–Osservatorio Astronomico di Padova, Vicolo dell’Osservatorio 5, I-35122, Padova, Italy}
\affiliation{Dipartimento di Fisica e Astronomia ``G.Galilei,'' Universitá di Padova, Via Marzolo 8, I-35131 Padova, Italy}
\author[0000-0003-3441-903X]{V{\'e}ronique Buat}
\affiliation{Aix Marseille Université, CNRS, CNES, LAM, Marseille, France}
\author[0000-0002-6184-9097]{Jaclyn B. Champagne}
\affiliation{Steward Observatory, University of Arizona, 933 N Cherry Ave, Tucson, AZ 85721, USA}
\author[0000-0003-3881-1397]{Olivia Cooper}
\altaffiliation{NSF Graduate Research Fellow}
\affiliation{Department of Astronomy, The University of Texas at Austin, 2515 Speedway Blvd Stop C1400, Austin, TX 78712, USA}
\author[0000-0001-6820-0015]{Luca Costantin}
\affiliation{Centro de Astrobiología (CAB/CSIC-INTA), Ctra. de Ajalvir km 4, Torrej{\'o}n de Ardoz, E-28850 Madrid, Spain}
\author[0000-0002-6219-5558]{Alexander de la Vega}
\affiliation{Department of Physics and Astronomy, University of California Riverside, 900 University Ave, Riverside, CA 92521, USA}
\author[0000-0003-4761-2197]{Nicole E. Drakos}
\affiliation{Department of Astronomy and Astrophysics, University of California, Santa Cruz, 1156 High Street, Santa Cruz, CA 95064, USA}
\author[0000-0002-9382-9832]{Andreas Faisst}
\affiliation{Caltech/IPAC, 1200 E. California Blvd., Pasadena, CA 91125, USA}
\author[0000-0003-3820-2823]{Adriano Fontana}
\affiliation{INAF, Osservatorio Astronomico di Roma, via di Frascati 33, 00078 Monte Porzio Catone, Italy}
\author[0000-0002-7530-8857]{Seiji Fujimoto}
\altaffiliation{Hubble Fellow}
\affiliation{Department of Astronomy, The University of Texas at Austin, 2515 Speedway Blvd Stop C1400, Austin, TX 78712, USA}
\author[0000-0001-9885-4589]{Steven Gillman}
\affiliation{Cosmic Dawn Center (DAWN), Denmark}
\affiliation{DTU-Space, Technical University of Denmark, Elektrovej 327, DK-2800 Kgs. Lyngby, Denmark}
\author[0000-0002-4085-9165]{Carlos G{\'o}mez-Guijarro}
\affiliation{Universit{\'e} Paris-Saclay, Universit{\'e} Paris Cit{\'e}, CEA, CNRS, AIM, 91191, Gif-sur-Yvette, France}
\author[0000-0002-0236-919X]{Ghassem Gozaliasl}
\affiliation{Department of Physics, University of Helsinki, P.O. Box 64, FI-00014 Helsinki, Finland}
\author[0000-0001-6145-5090]{Nimish P. Hathi}
\affiliation{Space Telescope Science Institute, 3700 San Martin Drive, Baltimore, MD 21218, USA}
\author[0000-0003-4073-3236]{Christopher C. Hayward}
\affiliation{Center for Computational Astrophysics, Flatiron Institute, 162 Fifth Avenue, New York, NY 10010, USA}
\author[0000-0002-3301-3321]{Michaela Hirschmann}
\affiliation{Institute for Physics, Laboratory for Galaxy Evolution and Spectral modelling, Ecole Polytechnique Federale de Lausanne, Observatoire de Sauverny, Chemin Pegasi 51, 1290 Versoix, Switzerland}
\affiliation{INAF, Osservatorio Astronomico di Trieste, Via Tiepolo 11, 34131 Trieste, Italy}
\author[0000-0002-4884-6756]{Benne W. Holwerda}
\affiliation{Physics \& Astronomy Department, University of Louisville, 40292 KY, Louisville, USA}
\author[0000-0002-8412-7951]{Shuowen Jin}
\affiliation{Cosmic Dawn Center (DAWN), Denmark}
\affiliation{DTU-Space, Technical University of Denmark, Elektrovej 327, DK-2800 Kgs. Lyngby, Denmark}
\author[0000-0002-8360-3880]{Dale D. Kocevski}
\affiliation{Department of Physics and Astronomy, Colby College, Waterville, ME 04901, USA}
\author[0000-0002-5588-9156]{Vasily Kokorev}
\affiliation{Kapteyn Astronomical Institute, University of Groningen, P.O. Box 800, 9700 AV Groningen, The Netherlands}
\author[0000-0003-3216-7190]{Erini Lambrides}
\affiliation{NASA Goddard Space Flight Center, 8800 Greenbelt Rd, Greenbelt, MD 20771, USA}
\altaffiliation{NPP Fellow}
\author[0000-0003-1581-7825]{Ray A. Lucas}
\affiliation{Space Telescope Science Institute, 3700 San Martin Drive, Baltimore, MD 21218, USA}
\author[0000-0002-4872-2294]{Georgios E. Magdis}
\affiliation{Cosmic Dawn Center (DAWN), Denmark}
\affiliation{Niels Bohr Institute, University of Copenhagen, Jagtvej 128, DK-2200, Copenhagen N, Denmark}
\affiliation{DTU-Space, Technical University of Denmark, Elektrovej 327, DK-2800 Kgs. Lyngby, Denmark}
\author[0000-0002-6777-6490]{Benjamin Magnelli}
\affiliation{Universit{\'e} Paris-Saclay, Universit{\'e} Paris Cit{\'e}, CEA, CNRS, AIM, 91191, Gif-sur-Yvette, France}
\author[0000-0002-6149-8178]{Jed McKinney}
\affiliation{Department of Astronomy, The University of Texas at Austin, 2515 Speedway Blvd Stop C1400, Austin, TX 78712, USA}
\author[0000-0001-5846-4404]{Bahram Mobasher}
\affiliation{Department of Physics and Astronomy, University of California Riverside, 900 University Ave, Riverside, CA 92521, USA}
\author[0000-0003-4528-5639]{Pablo G. P{\'e}rez-Gonz{\'a}lez}
\affiliation{Centro de Astrobiología (CAB/CSIC-INTA), Ctra. de Ajalvir km 4, Torrej{\'o}n de Ardoz, E-28850 Madrid, Spain}
\author[0000-0003-0427-8387]{R. Michael Rich}
\affiliation{Department of Physics and Astronomy, UCLA, PAB 430 Portola Plaza, Box 951547, Los Angeles, CA 90095-1547}
\author[0000-0001-7755-4755]{Lise-Marie Seill{\'e}}
\affiliation{Aix Marseille Université, CNRS, CNES, LAM, Marseille, France}
\author[0000-0003-4352-2063]{Margherita Talia}
\affiliation{University of Bologna - Department of Physics and Astronomy “Augusto Righi” (DIFA), Via Gobetti 93/2, I-40129 Bologna, Italy}
\affiliation{INAF, Osservatorio di Astrofisica e Scienza dello Spazio, Via Gobetti 93/3, I-40129, Bologna, Italy}
\author[0000-0002-0745-9792]{{C. Megan} {Urry}}
\affiliation{Physics Department and Yale Center for Astronomy \& Astrophysics, Yale University, PO Box 208120, CT 06520, USA}
\author[0000-0001-6477-4011]{Francesco Valentino}
\affiliation{European Southern Observatory, Karl-Schwarzschild-Strasse 2, D-85748, Garching bei München, Germany}
\affiliation{Cosmic Dawn Center (DAWN), Denmark}
\author[0000-0001-7160-3632]{Katherine E. Whitaker}
\affiliation{Department of Astronomy, University of Massachusetts, Amherst, MA 01003, USA}
\affiliation{Cosmic Dawn Center (DAWN), Denmark}
\author[0000-0003-3466-035X]{{L. Y. Aaron} {Yung}}
\affiliation{NASA Goddard Space Flight Center, 8800 Greenbelt Rd, Greenbelt, MD 20771, USA}
\altaffiliation{NASA Postdoctoral Fellow}
\author[0000-0002-7051-1100]{Jorge Zavala} 
\affiliation{National Astronomical Observatory of Japan, 2-21-1 Osawa, Mitaka, Tokyo 181-8588, Japan}
\collaboration{59}{and the COSMOS-Web and CEERS teams \vspace{-20pt}}

\submitjournal{(Affiliations can be found after the references)}

\begin{abstract} 
We present a search for extremely red, dust-obscured, $z>7$ galaxies with \textit{JWST}/NIRCam+MIRI imaging over the first 20 arcmin$^2$ of publicly-available Cycle 1 data from the COSMOS-Web, CEERS, and PRIMER surveys. 
Based on their red color in F277W$-$F444W ($\sim 2.5$\,mag) and detection in MIRI/F770W ($\sim$25\,mag), we identify two galaxies---\COS\ and \CEERS---which have best-fit photometric redshifts of $z=8.5^{+0.3}_{-0.4}$ and $z=7.6^{+0.1}_{-0.1}$, respectively. 
We perform SED fitting with a variety of codes (including \bagpipes, \prospector, \beagle, and \cigale), and find a $>95\%$ probability that these indeed lie at $z>7$.
Both sources are compact ($R_{\rm eff} \lesssim 200$ pc), highly obscured ($A_V \sim 1.5$--$2.5$), and, at our best-fit redshift estimates, likely have strong $[{\rm O}\,\textsc{iii}]$+H$\beta$ emission contributing to their $4.4\,\mu$m photometry.
We estimate stellar masses of $\sim 10^{10}~M_\odot$ for both sources; by virtue of detection in MIRI at $7.7\,\mu$m, these measurements are robust to the inclusion of bright emission lines, for example,~from an AGN. 
We identify a marginal (2.9$\sigma$) ALMA detection at 2 mm within $0.5$''\,of \COS, which if real, would suggest a remarkably high IR luminosity of $\sim 10^{12} L_\odot$. 
These two galaxies, if confirmed at $z\sim 8$, would be extreme in their stellar and dust masses, and may be representative of a substantial population of modestly dust-obscured galaxies at cosmic dawn. 
\end{abstract}

\section{Introduction}\label{sec:intro}

The launch of \textit{JWST} has immensely widened our view of the $z>8$ Universe, pushing observations within a mere 600 Myr of the Big Bang. 
Within the first year of observations, dozens of candidate $z\sim8$--$10$ galaxies have been identified based on photometric redshifts \citep{naiduTwo2022,finkelsteinCEERS2022b, donnanEvolution2023a}, and several have been spectroscopically confirmed with \textit{JWST}/NIRSpec \citep[e.g.][]{fujimotoCEERS2023b, larsonCEERS2023, saxenaJADES2023, curtis-lakeSpectroscopic2022, arrabalharoSpectroscopic2023, tangJWST2023}. 
Virtually all of these candidates have been selected via the Lyman break \citep{steidelSpectroscopic1996}, a technique which is most effective for intrinsically blue and UV-luminous sources but fails to capture fainter, reddened objects.

Galaxy selection via the Lyman break in the rest-frame UV poses a particular challenge for measuring accurate stellar masses \citep[e.g.][]{papovichCEERS2022} and placing the candidates in a cosmological context. 
The most massive galaxies in the early Universe can provide key constraints on physical models of galaxy formation and evolution \citep[e.g.][]{boylan-kolchinStress2022a, harikaneComprehensive2023a, ferraraStunning2022, menciHighredshift2022, lovellExtreme2023}, and will by nature be more chemically evolved and may have substantial dust reservoirs \citep{whitakerConstant2017a}. 
Indeed, a strong candidate population for the most massive galaxies in the early Universe are dusty star-forming galaxies (DSFGs), which are ubiquitous at cosmic noon ($z\sim 2$), where they dominate the star-formation rate density (SFRD) of the Universe \citep*{caseyDusty2014, madauCosmic2014}. 
While DSFGs have been detected out to $z\sim 6$--$7$, with obscured star-formation rates in excess of $1000$ M$_\odot$ yr$^{-1}$  \citep{zavalaDusty2018,marroneGalaxy2018,caseyPhysical2019,endsleyALMA2022,fujimotoDusty2022}, their volume density at this epoch remains largely unconstrained due to the difficulty of constructing a complete sample. 
In particular, accurate photometric redshift estimates and systematic spectroscopic follow-up for faint DSFGs is challenging due to their faint rest-UV and optical  emission \citep[e.g.][]{taliaIlluminating2021} and degeneracies with lower-redshift galaxies \citep[which are $\sim 20-100$ times more numerous, e.g.][]{smolcicMillimeter2012}.

Various techniques have been developed to constrain the population of dusty galaxies at $z\gtrsim 4$--$5$, including focusing on gravitationally lensed objects \citep[e.g.][]{zavalaDusty2018, marroneGalaxy2018, fujimotoALMA2021} or observing at longer wavelengths (e.g.~2--3 mm) to efficiently filter-out lower-redshift galaxies \citep[e.g.][]{caseyMapping2021, zavalaEvolution2021, cooperSearching2022}. 
However, at the highest redshifts ($z>7$), these methods have thus far only proven sensitive to the brightest, most extreme ($L_{\rm IR} \gtrsim 10^{12\mbox{--}13}\,L_\odot$) DSFGs, as confirming lower-luminosity sources at the highest redshifts is like searching for a ``needle in a haystack,'' with a relatively low yield at $z>4$ \citep{caseyAnalysis2018,caseyBrightest2018a}.
At present, most $z>7$ dust-continuum detections come from follow-up of UV-selected LBGs \citep[e.g.][]{watsonDusty2015, bouwensEvolution2022, fujimotoDusty2022} or serendipitous detections \citep[e.g.][]{fudamotoNormal2021} with only one robust dust-continuum-detected object at $z>8$: MACS0416-Y1 at $z=8.31$ \citep{tamuraDetection2019}.\footnote{A2744-YD4 had previously been reported at $z=8.38$ \citep{laporteDust2017}, but new \JWST\ spectroscopic observations confirm its redshift as $z=7.88$ \citep{morishitaEarly2022}.}
It remains unclear whether the perceived rarity of DSFGs in the first $\sim 750$ Myr of the Universe is intrinsic, due to a gradual buildup of dust/stellar mass, or artificial, due to the difficulty of detecting and confirming these objects.

For the first time, the unprecedented sensitivity and continuous infrared wavelength coverage of \textit{JWST} makes it possible to detect the rest-frame optical emission from extremely obscured galaxies at $z>8$. 
While far-infrared constraints at this epoch are limited, the near and mid-IR has the potential to identify the most massive, obscured galaxies based on the reddened stellar continuum, rest-frame Balmer break, and contributions from bright emission lines \cite[e.g.][]{labbePopulation2023, perez-gonzalezCEERS2022, rodighieroJWST2023}.
In particular, the mid-infrared coverage of \textit{JWST}/MIRI allows constraints on the full rest-frame optical SED out to $z\sim 9$. 
A unique advantage of long wavelength MIRI imaging comes in identifying and characterizing the earliest dust-obscured galaxies, which are most heavily obscured at rest-frame UV wavelengths. 
These objects will be critical to our understanding of the assembly of massive galaxies \citep{narayananFormation2015, longMissing2022} as well as the physical processes responsible for the buildup of large dust reservoirs, whether it be from asymptotic giant branch (AGB) stars, supernovae (SNe), or efficient grain growth in the interstellar medium \citep{dwekOrigin2011, jonesEvolution2013, michalowskiDust2015, lesniewskaDust2019}.

In this paper, we present a search for extremely red, $z\gtrsim 7$ galaxies across the overlapping NIRCam+MIRI coverage in three Cycle 1 Treasury surveys: COSMOS-Web \citep{caseyCOSMOSWeb2022b}, CEERS \citep{finkelsteinCEERS2022b}, and PRIMER-COSMOS (P.I.~J.~Dunlop, GO\#1837). 
We report the detection of two galaxies, in CEERS and COSMOS-Web, with remarkably red colors and photometric redshifts at $z\sim 8$. 
Both are among the reddest galaxies identified in the entirety of their respective surveys, and detected in $7.7\,\mu$m MIRI imaging which provides robust constraints on their stellar mass.

The paper is organized as follows. 
In Section~\ref{sec:data} we describe an overview of the imaging data used and the construction of photometric catalogs. 
In Section~\ref{sec:selection} we describe our sample selection and vetting of individual candidates. 
In Section~\ref{sec:results} we present and characterize the two robust candidates, including their photometric redshifts (\S\ref{sec:photozs}), sizes (\S\ref{sec:size}), stellar masses (\S\ref{sec:mstar}), and far-infrared emission (\S\ref{sec:FIR}). 
Finally, in Section~\ref{sec:discussion} we discuss the implications of these discoveries for galaxy formation within a $\Lambda{\rm CDM}$ framework. 
Throughout this work, we assume a \textit{Planck} cosmology \citep{planckcollaborationPlanck2020} and a \citet{kroupaInitial2002a} stellar initial mass function (IMF). 
All quoted magnitudes are in the AB system \citep{okeAbsolute1974}.

\section{Data}\label{sec:data}

We utilize $1$--$8~\mu$m \textit{JWST}/NIRCam+MIRI imaging from three publicly available Cycle 1 programs to identify candidate massive, dusty galaxies at $z\gtrsim 8$. 
Table~\ref{tab:depths} provides the approximate $5\sigma$ depth for point sources and the effective area (i.e.~the NIRCam+MIRI overlapping area) for each survey. 

\subsection{JWST Observations \& Data Reduction}
\subsubsection{COSMOS-Web}

COSMOS-Web is a large Cycle 1 treasury program imaging a contiguous $0.54$ deg$^2$ in the COSMOS field with NIRCam and $0.2$ deg$^2$ with MIRI in parallel \citep{caseyCOSMOSWeb2022b}. 
As of this writing, 6 of the 152 visits have been completed during observations executed in January 2023, constituting a contiguous 77~arcmin$^2$ ($\sim 4\%$ of the overall area), with 8.7~arcmin$^2$ of overlap between the NIRCam and MIRI coverage. 
The COSMOS-Web imaging includes four NIRCam filters---F115W, F150W, F277W, and F444W---and one MIRI filter, F770W, at an approximate $5\sigma$ depth of 26 AB mag.

The full details on the NIRCam and MIRI reduction process will be presented in upcoming papers (M.~Franco et al.~\textit{in prep} and S.~Harish et al.~\textit{in prep}, respectively) but are briefly described here. 
The raw NIRCam imaging was reduced by the \JWST\ Calibration Pipeline version 1.8.3, with the addition of several custom modifications \citep[as has also been done for other JWST studies, e.g.][]{finkelsteinCEERS2022b}, including the subtraction of $1/f$ noise and sky background. We use the Calibration Reference Data System (CRDS)\footnote{\url{jwst-crds.stsci.edu}} pmap 0989  which corresponds to NIRCam instrument mapping imap 0232.
The final mosaics are created in Stage 3 of the pipeline with a pixel size of $0\farcs03$/pixel. 
Astrometric calibration is conducted via the JWST \textsc{tweakreg} procedure, with a reference catalog based on a \textit{HST}/F814W $0\farcs03$/pixel mosaic in the COSMOS field with astrometry tied to Gaia-EDR3 \citep{gaiacollaborationGaia2018}. 
The median offset in RA and Dec between our reference catalog and the NIRCam mosaic is less than 5 mas.
The MIRI/F770W observations were reduced using version 1.8.4 of the \JWST\ Calibration pipeline, along with additional steps for background subtraction that was necessary to mitigate the instrumental effects. The resulting mosaic was resampled onto a common output grid with a pixel scale of 0\farcs06/pixel and aligned with ancillary \HST/F814W imaging of the region.

\subsubsection{CEERS}

The Cosmic Evolution Early Release Science survey (CEERS) is one of 13 early release science surveys designed to obtain and release reduced data in early Cycle 1. 
CEERS consists of a mosaic of 10 NIRCam and 9 MIRI pointings in the CANDELS Extended Groth Strip (EGS) field, alongside spectroscopy with NIRSpec and NIRCam WFSS. 
Each CEERS/NIRCam pointing includes 7 filters: F115W, F150W, F200W, F277W, F356W, F410M, and F444W. 
The MIRI pointings include a range of filters from F560W to F2100W; here we only use MIRI pointings 3, 6, 7, and 9, as the other pointings either have no NIRCam overlap (1 and 2) or do not include F770W (5 and 8), which is essential for the robust selection of our $z>7$ dusty galaxy candidates. 
These four pointings provide 7.8 arcmin$^2$ of overlap between MIRI/F770W and NIRCam imaging, at an approximate $5\sigma$ depth in F770W of $\sim 27$ mag. 
We utilize the NIRCam and MIRI reductions produced by the CEERS team. 
The NIRCam reduction is described in detail in \citet{bagleyCEERS2022}, and the MIRI reduction will be described in Yang et al.~(\textit{in prep}). 

\subsubsection{PRIMER-COSMOS}

The Public Release Imaging for Extragalactic Research (PRIMER) survey (P.I.~J.~Dunlop, GO\#1837) is a large Cycle 1 Treasury Program to image two HST CANDELS Legacy Fields (COSMOS and UDS) with NIRCam+MIRI. 
PRIMER is conducted with MIRI as the prime instrument, and NIRCam in parallel, with observations split between two windows with opposite observational position angles.
This configuration maximizes the overlap between the MIRI and NIRCam coverage, though at the time of this writing, only the first epoch has been observed, constituting just $4.1$ arcmin$^2$ of overlap between NIRCam+MIRI. 
PRIMER imaging includes 8 NIRCam bands (equivalent to CEERS plus F090W), plus two MIRI bands (F770W and F1800W). 
As part of the reduction of COSMOS-Web data, the COSMOS-Web team has conducted an independent processing of the PRIMER data in the COSMOS field and thus we include it in the analysis here. 
We note that while the PRIMER-COSMOS field is contained entirely within the COSMOS-Web footprint, there will not be significant overlap between the two surveys until the completion of COSMOS-Web in January 2024.

\subsection{Archival ground-based and HST data}

In addition to the new \JWST\ data, we utilize the existing multiwavelength imaging in the COSMOS and EGS fields.
For COSMOS (encompassing PRIMER-COSMOS) we include the $grizy$ imaging from Subaru/Hyper Suprime-Cam  \citep[HSC;][]{aiharaSecond2019}, and $YJHK_s$ imaging from the UltraVISTA survey \citep{mccrackenUltraVISTA2012}.  
We additionally utilize the \HST/ACS F814W imaging covering the entire COSMOS field \citep{koekemoerCOSMOS2007} and \textit{Spitzer}/IRAC imaging from the Cosmic Dawn Survey \citep{euclidcollaborationEuclid2022}. 
This is the same data as is used in \citet{weaverCOSMOS20202022} with the exception of the UltraVISTA data which is updated to the newest public data release 5.

For CEERS, we include the \HST/ACS and WFC3 imaging in F606W, F814W, F125W, and F160W from the CANDELS survey \citep{groginCANDELS2011a, koekemoerCANDELS2011}. 

\begin{deluxetable}{lcccc}
\setlength{\tabcolsep}{5pt}
\tablecaption{Combined NIRCam+MIRI area and approximate 5$\sigma$ depths for the three surveys included in this work. \label{tab:depths}}
\tablehead{
	\colhead{\multirow{2}{*}{Survey}} & \colhead{Area} & \colhead{F277W} & \colhead{F444W} & \colhead{F770W}\\[-0.7em]
	& \colhead{(arcmin$^2$)} & \colhead{($5\sigma$)}& \colhead{($5\sigma$)}& \colhead{($5\sigma$)}\\[-1.5em]}
\startdata
	COSMOS-Web$^\dagger$ & 8.7 & 28.3 & 28.2 & 26.0 \\
	CEERS & 7.8 & 29.2 & 28.6 & 27.1 \\
	PRIMER-COSMOS$^\dagger$ & 4.1 & 28.9 & 28.7 & 26.0 
\enddata
\tablerefs{COSMOS-Web \citep[][GO\#1727]{caseyCOSMOSWeb2022b}, CEERS \citep[][DD-ERS\#1345]{finkelsteinCEERS2022b}, PRIMER (P.I.~J.~Dunlop; GO\#1837)}
\tablecomments{Depths are computed in $0.3''$ ($0.6''$) diameter apertures for NIRCam (MIRI).}
\tablenotetext{\dagger}{Table reflects data available as of March 2023. Total NIRCam+MIRI overlap upon survey completion will be $650$ arcmin$^2$ for COSMOS-Web and $140$ arcmin$^2$ for PRIMER-COSMOS.}
\end{deluxetable}

\subsection{Multiwavelength Catalogs}\label{sec:photometry}

\subsubsection{COSMOS-Web \& PRIMER-COSMOS}
For both COSMOS-Web and PRIMER-COSMOS we conduct source detection, perform model-based  photometry, and construct multi-band catalogs using \texttt{SourceXtractor++} \citep[][hereafter \texttt{SE++}]{bertinSourceXtractor2020,kummelWorking2020}, an updated version of the popular \texttt{SExtractor} package \citep{bertinSExtractor1996}.
The use of \texttt{SE++} is motivated by the desire to make full use of the depth and filter coverage of seeing-limited ground-based data in COSMOS as well as high-resolution near-infrared \JWST\ imaging. 
We perform source detection on a $\chi^2$ detection image constructed from all four NIRCam bands; priors for the source centroid positions are determined from this detection image. 
For each detected source in the $\chi^2$ image, \texttt{SE++} then fits a Sérsic model convolved with the filter-specific PSF in each of the measurement band. 
Here we use model PSFs from \texttt{WebbPSF} \citep{perrinSimulating2012, perrinUpdated2014}.
The Sérsic model parameters (centroid position, $n$, $R_{\rm eff}$, $b/a$) are fit jointly between all bands, weighted by their respective S/N, while the total flux is fit independently in each band.
Details on the \texttt{SE++} catalogs for COSMOS-Web will be presented in M.~Shuntov et al.~(\textit{in prep}.).

Computing photometric uncertainties in model-based photometry is not trivial for sources that are undetected in a given band, where the \texttt{SE++} model is below the noise.
This can occasionally result in significantly underestimated errors in the dropout bands, where the source is not detected; as such, we set a noise floor for each band equal to the rms measured in circular apertures with a diameter of 0.3'' (for ACS/NIRCam), 0.6'' (for MIRI), and 2'' (for ground-based data). 
We adopt the measured depths from \citet{weaverCOSMOS20202022} and \citet{caseyCOSMOSWeb2022b}; we report depths in relevant \JWST\ filters in Table~\ref{tab:depths}. 
The \texttt{SE++} catalogs yield 1789 (585) objects detected with both NIRCam and MIRI for COSMOS-Web (PRIMER-COSMOS).

\subsubsection{CEERS}
For CEERS, we adopt the multi-band \texttt{SExtractor} catalog previously described in \citet{finkelsteinCEERS2022b}. 
Source detection for this catalog is done on an inverse-variance-weighted sum of the F277W and F356W images. 
The catalog includes all available \textit{HST}/ACS and WFC3 data and all \textit{JWST}/NIRCam bands, but not MIRI photometry. 
We therefore perform source detection on the MIRI F770W images using \texttt{astropy}/\texttt{photutils}. 
Before performing detection we convolve the image with a $5\times 5$ pixel Gaussian smoothing kernel with FWHM of 2 pixels in order to better identify faint objects. 
We set a detection threshold at 1.1 times the background RMS and a minimum area of 4 pixels. 
In order to be consistent with the NIRCam catalog, we compute photometry on the MIRI F560W and F770W images in small elliptical apertures using the \texttt{SourceCatalog} task in \texttt{photutils}. 
We use a Kron factor of 1.1 to restrict the aperture to the central region of each galaxy, maximizing the signal-to-noise, and apply a correction based on the median ratio of the flux in these small apertures to equivalent apertures using a Kron factor of 2.5. 
Similar to the NIRCam photometry \citep{finkelsteinCEERS2022b}, we find a correction of $\sim 1.5$. 
We then cross-match the catalog of MIRI-detected sources, based on the measured centroid positions, with the NIRCam catalog. 
This procedure yields 1190 objects detected with both NIRCam and MIRI.

\subsection{ePSF construction}

We construct empirical point spread functions (ePSFs) in each band by stacking bright stars in the field. 
The ePSF construction for CEERS is described in \citet{finkelsteinCEERS2022b}; we adopt the same PSFs used in that work. 
For COSMOS-Web, we construct ePSFs from a catalog of bright stars in the COSMOS field \citep[as described in][]{weaverCOSMOS20202022}. 
We inspect cutouts at the position of each star and remove saturated stars and compact galaxies which were previously misidentified. 
We then construct ePSFs using the \texttt{EPSFBuilder} module in \texttt{astropy/photutils} \citep{bradleyAstropy2022}, which follows the prescriptions of \citet{andersonHigh2000}. 
The ePSFs are slightly broader than the corresponding \texttt{WebbPSF} models due to the nature of the mosaicking process.

\section{Sample selection \& vetting}\label{sec:selection}

We determine initial photometric redshift estimates for the entire NIRCam+MIRI samples across all three imaging surveys using EAzY \citep{brammerEAZY2008}. 
EAzY computes linear combinations of pre-defined templates to derive probability distribution functions (PDFs) for the redshift based on the $\chi^2$ of the templates. 
We fit to all available ground-based, \textit{HST}, NIRCam and MIRI photometry, based on the multi-wavelength catalogs described in Section~\ref{sec:photometry}. 
The template set we use includes the standard \verb|tweak_fsps_QSF_12_v3| set of 12 FSPS \citep{conroyPropagation2010} templates, as well as the 6 templates from \citet{larsonSpectral2022}. 
We allow the redshift to vary from 0 to 15 with a step size of $\Delta z = 0.01$. 
Though we refine the photometric redshifts for individual sources of interest later, this first-pass photo-$z$ run allows us to explore the relationship between observed-frame colors and redshift and outline the sample selection of massive, dusty, and red $z>7$ galaxies in the context of the full catalog.  

\begin{figure}
\centering
\includegraphics[width=\linewidth]{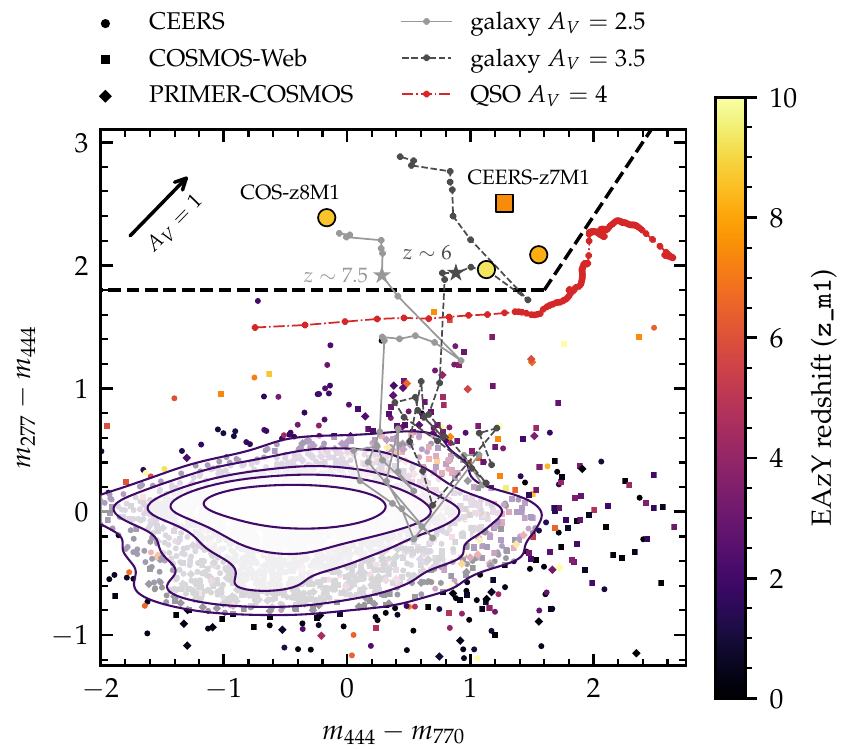}	
\caption{The $m_{277}-m_{444}$ vs.~$m_{444}-m_{770}$ color-color diagram indicating selection of $z\gtrsim 7$ dusty galaxies. Points are colored by their best-fit photometric redshift from first-pass EAzY runs. The dashed lines indicate our proposed color selection criteria, which captures 4 objects, all with photometric redshifts $\gtrsim 7$. 
We show in grey two model tracks generated from \bagpipes, for young (10 Myr) stellar populations with $A_V=2.5$ and $3.5$ from $z\sim 1$--$9$.
We additionally show in red a QSO model with $A_V=4$, which, at $z>3$, is redder in $m_{444}-m_{770}$ than the galaxy models.
The color-color selection shown here is optimized for selecting $z\gtrsim 6$--$7$ obscured galaxies with young ages and bright emission lines but rejecting objects with steeply rising MIR SEDs (e.g. obscured AGN).} \label{fig:colorcolor}
\end{figure}

\begin{figure*}[t!]
\centering
\includegraphics[width=\linewidth]{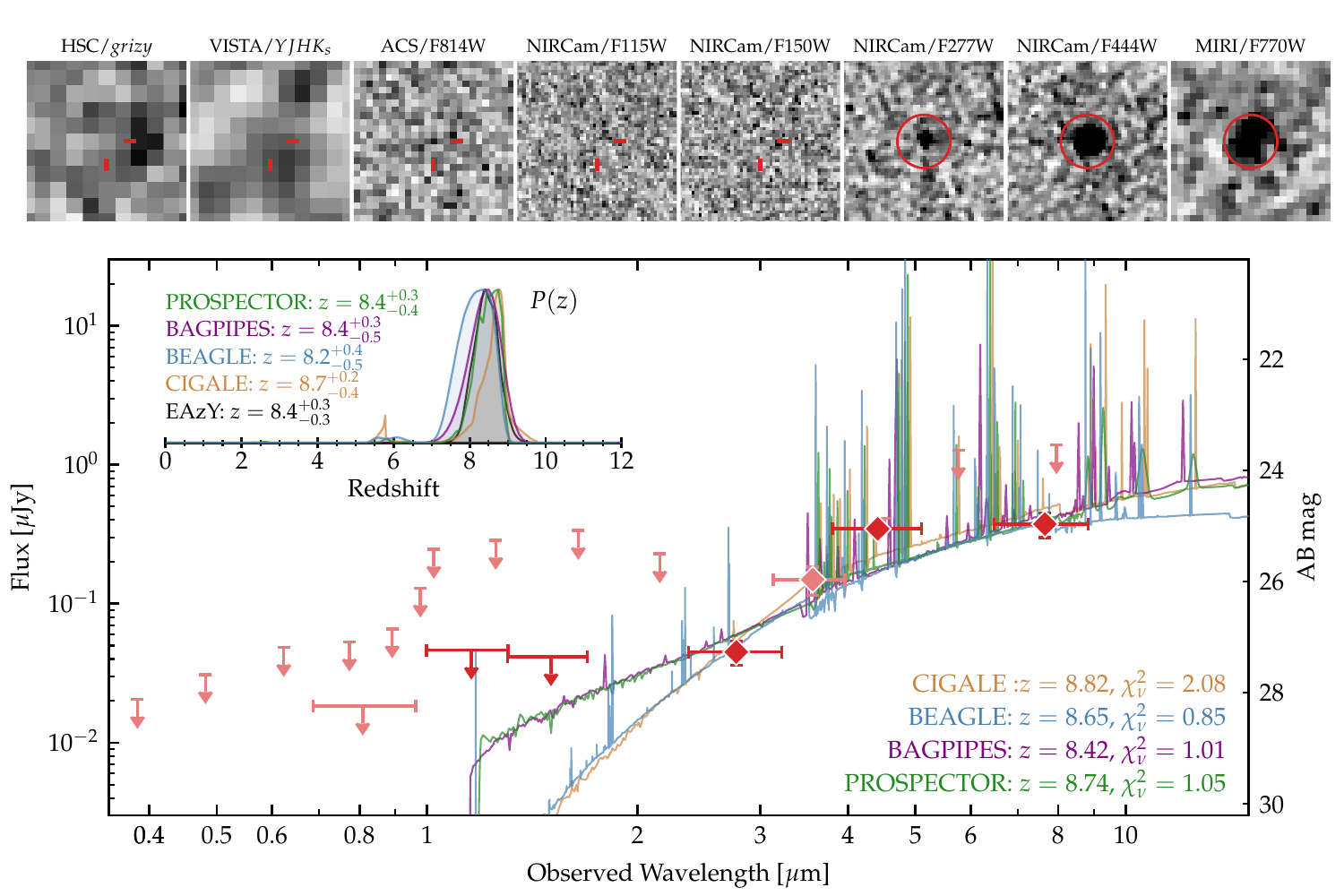}	
\caption{The optical through mid-infrared SED of \COS. The top panels show $1.5''$ square cutouts in the available \textit{HST}/JWST bands, plus stacked ground-based imaging from Subaru/HSC and UltraVISTA. The bottom panel shows the measured photometry (or 2$\sigma$ upper limits) from all available \JWST\ bands (shown in dark red) as well as \HST\ and ground-based bands (in light red). We additionally show maximum a posteriori (MAP) model SEDs from \cigale, \beagle, \bagpipes, and \prospector\ (for clarity we do not show the EAzY SED). The inset panel shows the full redshift probability distributions $P(z)$ from each photo-$z$ code. Robust detection in NIRCam/F444W and MIRI/F770W, with a sharp break between F444W and F277W, mandates a massive, dusty, $z>7$ solution.}\label{fig:sed1}
\end{figure*}

We identify high-redshift, dusty galaxy candidates based on their colors in $m_{444}-m_{770}$ and $m_{277}-m_{444}$. 
In particular, we design our selection to specifically target galaxies that are red in $m_{277}-m_{444}$ but not {\it as} red in $m_{444}-m_{770}$; this helps mitigate contamination from lower-redshift, extremely obscured sources or dusty AGN whose SED continues to rise in the mid-infrared due to hot dust in the AGN torus. 
Figure~\ref{fig:colorcolor} shows the color-color diagram for MIRI/F770W-detected galaxies in COSMOS-Web (circles), CEERS (squares), and PRIMER-COSMOS (diamonds). 
Specifically, we require $S/N_{770} > 5$, $S/N_{444} > 5$, and $S/N_{277} > 2$ to be included in our color-color plot and eventual sample. 
This ensures robust detection in F770W and F444W, and at least a marginal detection in F277W despite the red $m_{277}-m_{444}$ color. 
We additionally show in Figure~\ref{fig:colorcolor} two \bagpipes\ models (10 Myr old, 20\% solar metallicity) reddened by a Calzetti dust law with $A_V=2.5$ (solid line) and $A_V=3.5$ (dashed line). 
We also plot a QSO model by reddening the SDSS QSO composite spectrum \citep{vandenberkComposite2001, glikmanNearInfrared2006} with a Calzetti law with $A_V = 4$ from $1<z<8$.

We outline a rough color-color selection criteria to encompass the reddest galaxies in our sample, which all have EAzY photometric redshifts $>7$.
Specifically, we use
\begin{align}
	m_{277}-&m_{444} > 1.8 \qquad{\rm and} \\	
m_{277}-&m_{444} > 1.5(m_{444}-m_{770}) - 0.6
\end{align}
While the second criterion does not exclude any galaxies in this sample (i.e., this same selection could be done with just $m_{277}-m_{444}$), we include it to indicate that we require the $m_{444}-m_{770}$ color to be appreciably bluer than $m_{277}-m_{444}$, motivated by the need to reject objects with red mid-infrared SEDs, which are likely obscured AGN. 
Indeed, the reddened QSO template shown in Figure~\ref{fig:colorcolor} has $m_{277}-m_{444}>1.8$ at $z>2.5$, but is omitted from our selection criteria due to its redder color in $m_{444}-m_{770}$. 
These color-color criteria are optimized for selecting $z\gtrsim 6$--$7$ obscured galaxies with young ages and bright emission lines. 
Future work will refine this selection criteria with larger samples and more thorough modeling.

We find four galaxies satisfying our color-color selection based on the initial photometry: three in COSMOS-Web, one in CEERS, and none in PRIMER-COSMOS. 
In order to vet these candidates further we compute photometry in custom circular apertures on each galaxy. 
This is intended to reject spurious detections or underestimated uncertainties which may be present in the catalog photometry used in Figure~\ref{fig:colorcolor}. 
In particular, we use the \texttt{astropy}/\texttt{photutils} package to perform aperture photometry with aperture diameters ranging from $0.3$--$0.5$''. 
We correct the aperture flux based on the PSF flux falling outside the circular aperture.

With this photometry, we re-run EAzY and inspect the resulting solutions.  
Two of the four candidates have highly uncertain redshift probability distributions as estimated by EAzY; both are in COSMOS-Web.
We show the cutouts and SEDs for these two poorly-constrained candidates in Appendix~\ref{appendix:seds}, but for the remainder of the paper we focus on the two remaining candidates, which are significantly brighter and, therefore, much better constrained to $z>7$ with EAzY.

\begin{figure*}
\centering
\includegraphics[width=\linewidth]{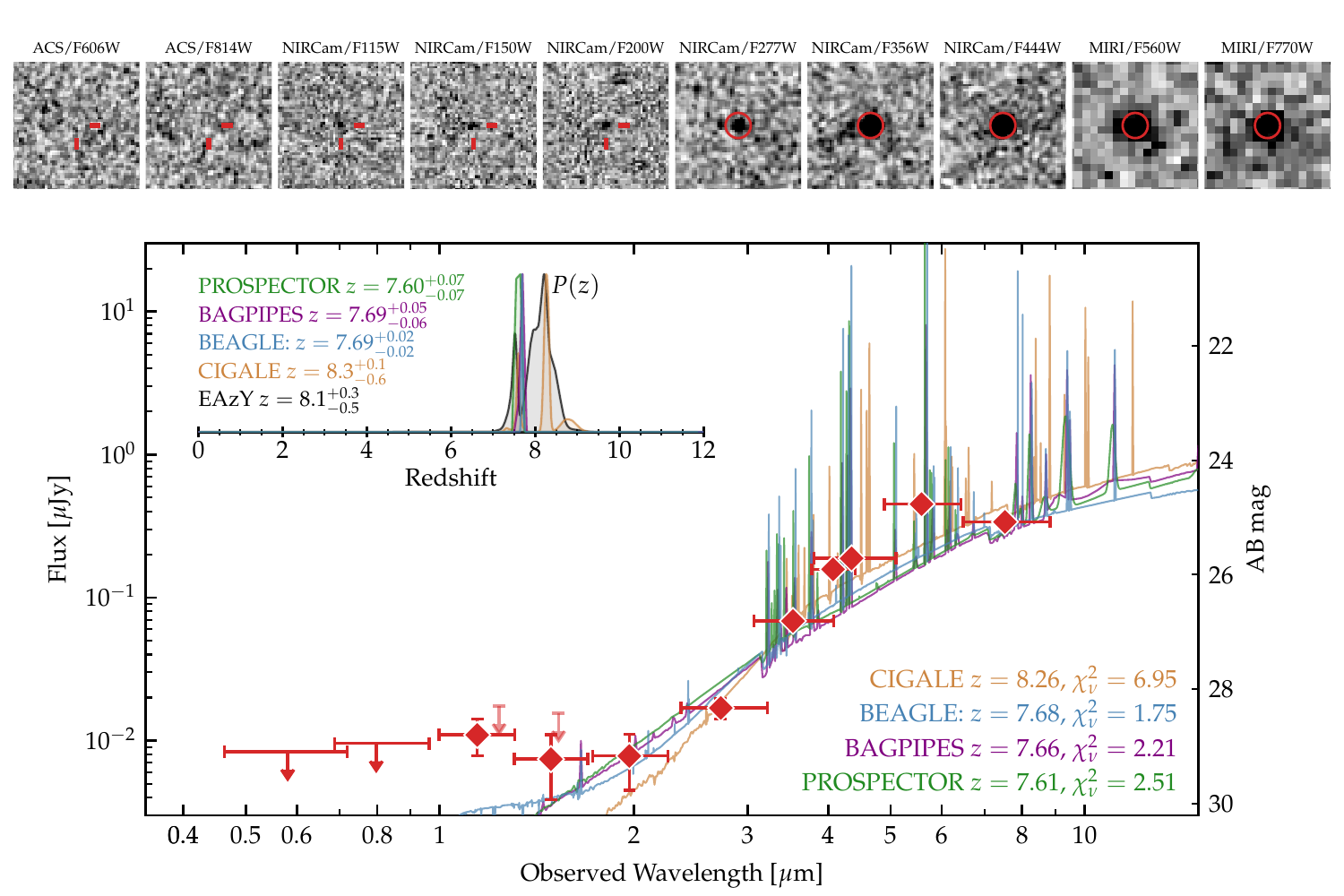}	
\caption{Same as Figure~\ref{fig:sed1}, but for \CEERS. The NIRCam SW bands reveal a blue UV slope ($\beta\sim -2.5$) at $1$--$2\,\mu$m despite the very red continuum at $>2\,\mu$m. 
This could be from SF in an unobscured line of sight, or an unobscured AGN component. 
The additional filters constrain the redshift precisely to $z=7.6$.}\label{fig:sed2}
\end{figure*}

To explore any possible low-redshift redundancy to our EAzY photometric redshift solutions, we explore fits with four other SED fitting codes: \prospector\ \citep{lejaDeriving2017a,johnsonStellar2021a}, \bagpipes\ \citep{carnallInferring2018a}, \cigale\ \citep{burgarellaStar2005a,nollAnalysis2009,boquienCIGALE2019}, and \beagle\ \citep{gutkinModelling2016,chevallardModelling2016}. 
For \bagpipes\ and \cigale\ we adopt a delayed-$\tau$ star-formation history (SFH) model, while for \beagle\ we adopt a constant SFH to be consistent with recent work \citep{endsleyJWST2022,whitlerAges2023,furtakJWST2022}. 
We additionally include, in all three codes, a late starburst (with age $\sim 10$--$100$ Myr) to allow for bright emission lines from \hii\ regions. 
We allow $A_V$ to vary from 0 to 4 and adopt a \citet{calzettiDust2000} attenuation law for \bagpipes, a \citet{charlotSimple2000b} law for \cigale, and the \citet{chevallardInsights2013} model for \beagle.  
For \prospector\ we use the \prospector-$\beta$ model, which includes a non-parametric star-formation history and informed, joint priors encoding empirical constraints on the redshifts, stellar masses, and SFHs of observed galaxies \citep[see][Table 1]{wangInferring2023}. 
Finally, we include nebular emission in all our fits. 
\beagle\ uses nebular emission templates from \citet{gutkinModelling2016} which combine the latest \citet{bruzualStellar2003a} stellar population models with \textsc{cloudy} nebular emission \citep{ferland20132013}. 
\bagpipes\ and \prospector\ implement nebular emission via updated \textsc{cloudy} models from \citet{bylerSelfconsistent2019a}, while \cigale\ uses an updated grid of \textsc{cloudy} models as described \citet{boquienCIGALE2019}.
While each of these fits use slightly different physical assumptions, the breadth of approaches here serves as a valuable test of the security of the candidates as genuine $z>7$ galaxies.

\section{Results}\label{sec:results}
\subsection{Two remarkable dust-obscured, $z\sim 8$ candidates}\label{sec:photozs}

Based on the sample selection and candidate vetting described in \S\ref{sec:selection}, we identify two robust $z\sim 8$ dusty galaxies, with very red colors ($m_{277}-m_{444} \sim 2.5$), which we denote as \COS\ and \CEERS\ for their selection via MIRI detections. 
Figures~\ref{fig:sed1} and \ref{fig:sed2} show the best-fit SEDs and redshift probability distributions for the two candidates.
We show $1.5''$ square cutouts in all available \HST/ACS, \JWST/NIRCam, and \JWST/MIRI bands, and additionally a stack of ground-based imaging for the COSMOS source. 
We plot the SEDs from the \bagpipes, \beagle, \cigale, and \prospector\ fits; for clarity we show the $P(z)$ from EAzY but not the best-fit SED. 
We note that the EAzY $\chi^2$ is similar to the other codes, and forcing a $z<7$ solution with EAzY yields a higher $\chi^2$ by $>10$. 

\vspace{-25pt}
\begin{deluxetable*}{lccccccc}
\setlength{\tabcolsep}{5pt}
\tablecaption{Physical properties derived from SED fitting for the two dust-obscured, $z\sim 8$ sources. \label{tab:properties}}
\colnumbers
\tablehead{
	\colhead{\multirow{2}{*}{ID}} & \colhead{R.A., Decl.} & \colhead{Photo-$z$} & \colhead{Photometric} & \colhead{$\log M_\star$} & \colhead{SFR$_{10\,\mathrm{Myr}}$}& \colhead{SFR$_{100\,\mathrm{Myr}}$} & \colhead{$A_V$}\\[-0.7em]
	& \colhead{[J2000]} &  \colhead{Code}&\colhead{Redshift} & \colhead{[${\rm M}_\odot$]} & \colhead{[${\rm M}_\odot~{\rm yr}^{-1}$]} & \colhead{[${\rm M}_\odot~{\rm yr}^{-1}$]} & \colhead{[mag]}\\[-2.5em]}
\startdata\\[-2em]
\multirow{5}{*}{\COS} & \multirow{5}{*}{\shortstack[c]{09h59m40.18s, \\ +02d17m32.19s}} & \textsc{prospector}-$\beta$ & $8.5^{+0.3}_{-0.4}$ & $9.8^{+0.1}_{-0.1}$ & $520^{+110}_{-100}$ & $59^{+22}_{-14}$ & $1.6^{+0.3}_{-0.2}$ \\
 &  &  \textsc{bagpipes} & $8.4^{+0.4}_{-0.5}$ & $9.9^{+0.2}_{-0.2}$ & $320^{+140}_{-90}$ & $73^{+51}_{-30}$ & $2.4^{+0.2}_{-0.2}$ \\
 &  & \textsc{beagle} & $8.2^{+0.4}_{-0.5}$ & 
$9.5^{+0.2}_{-0.2}$ & $270^{+120}_{-90}$ & $27^{+13}_{-8}$ & $1.9^{+0.3}_{-0.3}$ \\
&  &  \textsc{cigale} & $8.7^{+0.2}_{-0.4}$ & $10.3^{+0.4}_{-0.3}$ & $480^{+290}_{-460}$ & $57^{+35}_{-24}$ & $1.1^{+0.3}_{-0.1}$\\
&  &  \textsc{eazy} & $8.4^{+0.3}_{-0.3}$ & \dots & \dots & \dots & \dots \\[1mm]\hline\\[-4mm]
\multirow{5}{*}{\CEERS} & \multirow{5}{*}{\shortstack[c]{14h19m43.08s, \\ +52d53m16.48s}} & \textsc{prospector}-$\beta$ & $7.60^{+0.07}_{-0.07}$ & $10.1^{+0.1}_{-0.1}$ & $1010^{+270}_{-320}$ & $120^{+36}_{-33}$ & $2.6^{+0.1}_{-0.2}$ \\ 
 & & \textsc{bagpipes} & $7.69^{+0.05}_{-0.06}$ & $9.7^{+0.2}_{-0.2}$ & $360^{+100}_{-150}$ & $43^{+30}_{-20}$ & $3.2^{+0.2}_{-0.1}$ \\
 & & \textsc{beagle} & $7.69^{+0.02}_{-0.02}$ & $9.8^{+0.1}_{-0.1}$ & $600^{+120}_{-100}$ & $61^{+12}_{-10}$ & $3.1^{+0.1}_{-0.1}$ \\
 & & \textsc{cigale} & $8.26^{+0.12}_{-0.64}$ &  $10.2^{+0.3}_{-0.1}$ & $810^{+190}_{-270}$ & $86^{+21}_{-16}$ & $1.5^{+0.4}_{-0.1}$ \\
 & & \textsc{eazy} & $8.22^{+0.31}_{-0.70}$ & \dots & \dots & \dots & \dots 
\enddata
\tablecomments{Stellar masses and SFRs from \beagle\ and \cigale\ are corrected by a factor of 1.12 to be consistent with our assumption of a \citet{kroupaInitial2002a} IMF.\vspace{-0.7cm}}
\end{deluxetable*}

\COS\ (Figure~\ref{fig:sed1}), is only detected in F277W, F444W, F770W, and in marginally in \textit{Spitzer}/IRAC [3.6]. 
The extremely red $m_{277}-m_{444}$ color despite a relatively blue $m_{444}-m_{770}$ color drives the redshift solutions to $z>7$.
This SED shape could be due to the redshifted rest-frame $4000$ \AA\ break \citep[e.g.][]{labbePopulation2023}, which would suggest a maximally old stellar population which formed most of its mass by $z\sim 15$. 
Alternatively, the SED shape could be due to the contribution of \oiii\ 5007~\AA\ emission to the F444W flux. 
Bright emission lines have been shown to play a significant role in elevating broad-band photometry in high-redshift galaxies \citep[e.g.][]{shimHa2011, starkKeck2013, faisstCoherent2016, mckinneyBroad2022,fujimotoALMA2022a, naiduSchrodinger2022a, zavalaDusty2023,arrabalharoSpectroscopic2023}, and longer-wavelength MIRI data is critical to constrain underlying continuum \citep{papovichCEERS2022}.
Indeed, the emission-line dominated scenario is preferred by our SED fitting codes, as shown in Figure~\ref{fig:sed1}.
At $7<z<9$, \oiii\ 5007~\AA\ emission falls into F444W while H$\alpha$ falls blueward of F770W, yielding an extreme color differential. 
A high \oiii+H$\beta$ EW of $\sim 800$ \AA\ is needed to contribute sufficiently to the F444W flux. 
This is consistent with recent spectroscopic results for $z>7$ galaxies, for which the \oiii\ 5007~\AA\ emission line is known to be particularly bright \citep[]{fujimotoCEERS2023b,saxenaJADES2023, trumpPhysical2023}.

Similarly, for \CEERS, the redshift solution is significantly constrained by the elevated flux in NIRCam/F410M+F444W and MIRI/F560W, likely due to bright \oiii\ and H$\alpha$ emission, respectively. 
The source is well-detected in all NIRCam LW and MIRI bands (S/N$>10$), and the contribution from bright emission lines constrains the redshift to $z\sim 7.6$.
Importantly, the redshift probability distribution is consistent, albeit broader, if we fit only the COSMOS-Web filter set at the appropriate depth (e.g.~only F277W, F444W, and F770W). 
This is promising for the fidelity of finding similar objects in large surveys like COSMOS-Web.

We do note that \CEERS\ is marginally detected in NIRCam F115W, F150W, and F200W, which may suggest the presence of a less-obscured, UV-luminous component to the galaxy.
This is not inconsistent with the SED for \COS, for which the short wavelength data is too shallow to constrain any UV emission.
Since there is no signal in ACS/F606W or F814W, we consider the $z>7$ solution robust; indeed, fitting EAzY to only $\lambda <2\,\mu$m data gives $z\sim 6$--$8$. 
While none of the SED fitting codes capture this blue component, this is to be expected, as these codes assume a uniform dust screen attenuating all the starlight. In reality, some UV emission may escape unattenuated in the case of patchy dust.

Table~\ref{tab:properties} gives the physical properties derived from SED fitting for \COS\ and \CEERS, for each SED fitting code used. 
The different codes generally agree on high stellar mass for this epoch ($\log M_\star/{\rm M}_\odot > 9.5$), a steeply rising SFH (with SFR$_{10\,\mathrm{Myr}}$/SFR$_{100\,\mathrm{Myr}} \sim 10$), and a high dust attenuation ($A_V \sim 1.5$--$2.5$ mags) for both candidates. 
For the sake of simplicity, we adopt the photometric redshifts and physical parameters from \prospector\ for the remainder of this paper, though we add an additional $0.2$ dex uncertainty to the stellar mass to capture the differences between different fits. 
However, given that the different fits are consistent, despite differences in their assumptions for stellar population synthesis/dust attenuation and different sampling algorithms, these results are not particularly sensitive to the modeling assumptions.

\subsection{Rest-frame Optical Sizes}\label{sec:size}

We use the 2D image fitting code \textsc{imfit}\footnote{\url{https://www.mpe.mpg.de/~erwin/code/imfit/}}  \citep{erwinIMFIT2015} to characterize the rest-frame optical sizes of \COS\ and \CEERS.
In particular, we fit PSF-convolved models to the NIRCam/F444W images. 
While F444W has the largest PSF of all the NIRCam bands, it is also the highest S/N detection for both sources.
We use the F444W ePSF used in our photometry as described in Section~\ref{sec:photometry}.

\begin{figure}
\centering
\includegraphics[width=\linewidth]{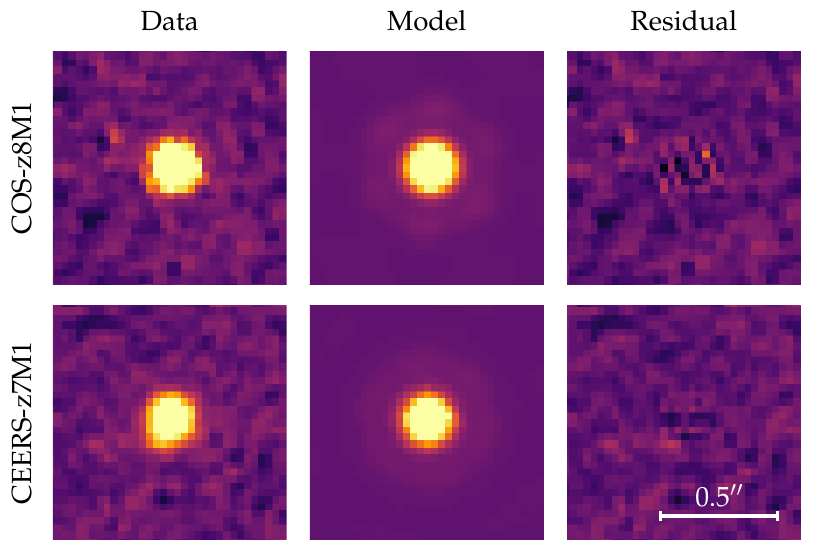}	
\caption{Results of 2D point-source profile fitting to the NIRCam/F444W imaging for both sources. The columns show the data, best-fit model, and residuals for each source. Both sources are well characterized by a point source model, suggesting extremely compact sizes. We derive an upper limit on the true sizes of $R_{\rm eff}\lesssim 200$ pc, as discussed in the text.\vspace{-0.5cm} }\label{fig:morphology}
\end{figure}

We find that both sources are well fit by a point-source model.
Figure~\ref{fig:morphology} shows the results of the point-source fitting for the two candidates (data, model, and residual). 
While fitting a Sérsic model yields a marginal improvement in the reduced $\chi^2$ statistic, the resulting Sérsic parameters are poorly constrained.
In order to provide a constraint on $R_{\rm eff}$, we fix the Sérsic index $n=1$, the axis ratio $b/a=1$ and the position angle $\theta=0$ (i.e.~an exponential disk profile). 
We fit a series of models with $R_{\rm eff} \sim 0\farcs01$--$0\farcs07$ ($50$--$350$ pc at $z=8$). 
We find that the resulting residuals are significant at the 3$\sigma$ level for $R_{\rm eff} \gtrsim 200$ pc, but consistent with the background rms for $R_{\rm eff} \lesssim 200$ pc, so we adopt this as an upper limit on the true size.

We note that such compactness is relatively common in high-redshift galaxies observed with \JWST\ \citep{robertsonDiscovery2022a,roberts-borsaniShot2022,onoMorphologies2022,tacchellaJADES2023}. 
Nonetheless, given that neither source is resolved, we discuss the possibility that the source's emission is dominated by AGN in Section~\ref{sec:agn}. 
However, we note that the derived stellar masses are likely robust to any contamination from strong emission lines from an AGN.

\subsection{Stellar masses}\label{sec:mstar}

\begin{figure}
\centering
\includegraphics[width=\linewidth]{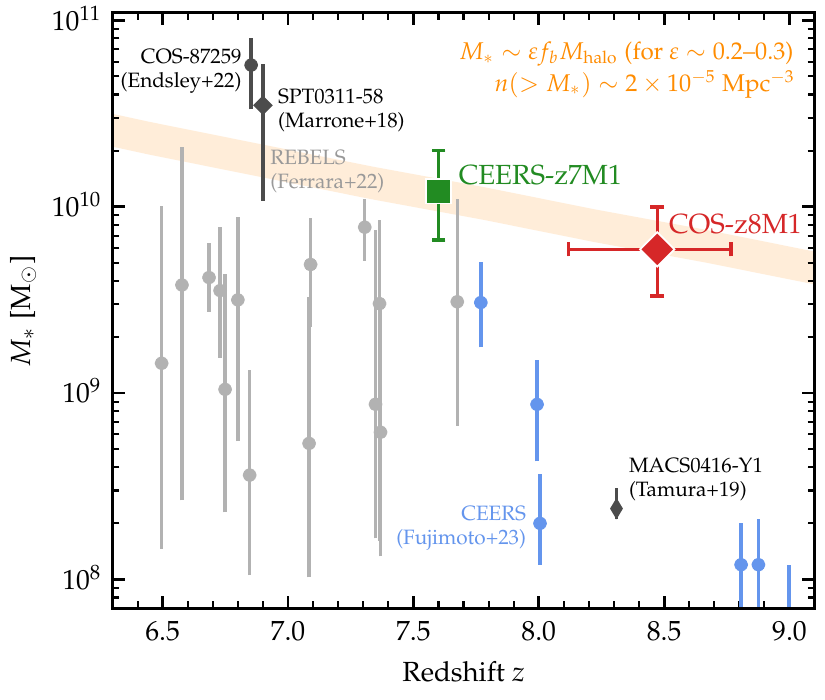}	
\caption{Inferred stellar mass vs.~redshift for \COS\ (red) and \CEERS\ (green), adopting the results from \prospector. We additionally show measurements for spectroscopically-confirmed, dust continuum-detected star-forming galaxies from the literature \citep{marroneGalaxy2018, tamuraDetection2019, endsleyALMA2022, ferraraALMA2022} as well as recent UV-bright galaxies confirmed with \JWST/NIRSpec \citep{fujimotoCEERS2023b}. The orange stripe shows the range of stellar masses expected for sources of this rarity, computed from an evolving halo mass function assuming a baryon conversion efficiency $\epsilon \sim 20$--$30\%$. Both \COS\ and \CEERS\ appear to be among the most massive dust-obscured galaxies at this epoch.}\label{fig:mstar_z}
\end{figure}

We compare the stellar masses derived from SED fitting for these two sources to estimates for similar objects in the literature. 
Figure~\ref{fig:mstar_z} shows the shows stellar mass vs.~redshift for \COS, \CEERS, and numerous dust-obscured, spectroscopically-confirmed galaxies in the literature \citep{marroneGalaxy2018,tamuraDetection2019,endsleyALMA2022,ferraraALMA2022}. 
We additionally show the sample of NIRCam-selected, unobscured, spectroscopically confirmed CEERS objects from \citet{fujimotoCEERS2023b}. 
Both \COS\ and \CEERS\ represent the extreme end of the dust-obscured high-$z$ population, with stellar masses $\gtrsim 3$ times higher than other known dust-continuum detected objects at the same redshift. 
We note that the various SED fitting codes used in this work all yield consistently large stellar masses, even when accounting for extremely high EW emission lines and differing dust attenuation laws (as discussed in Section~\ref{sec:photozs}).

The detection of these two sources at $z\sim 7$--$9$, across $20$ arcmin$^2$ of combined NIRCam+MIRI imaging, suggests an approximate volume density of $\sim 2\times 10^{-5}$ Mpc$^{-3}$. 
We estimate the typical stellar mass for sources of this rarity based on the evolving halo mass function. 
We compute the halo mass function using the python package \texttt{hmf} \citep{murrayHMFcalc2013}. We adopt a \citet{tinkerHalo2008} parametrization, modified with the redshift-dependent parameters from \citet{rodriguez-pueblaHalo2016} to be consistent with a \textit{Planck} cosmology \citep[see also discussion in][]{yungAre2023}. 
We compute the halo mass $M_{\rm halo}$ associated with a volume density $n(>M_{\rm halo}) \approx 2\times10^{-5}$ Mpc$^{-3}$. 
This halo mass is then converted to a stellar mass assuming the cosmic baryon fraction of $15.8\%$ and a baryon conversion efficiency, $\epsilon$. 
The orange stripe in Figure~\ref{fig:mstar_z} shows the expected stellar masses for $\epsilon \sim 20$--$30\%$; we don't include uncertainties (e.g.~from cosmic variance) into the width of the stripe, but rather just use it to illustrate the expected range of masses for different halo growth histories from $6.5<z<9$.

\subsection{Far-Infrared Constraints and Dust Mass}\label{sec:FIR}

\begin{figure}[t!]
\centering
\includegraphics[width=\linewidth]{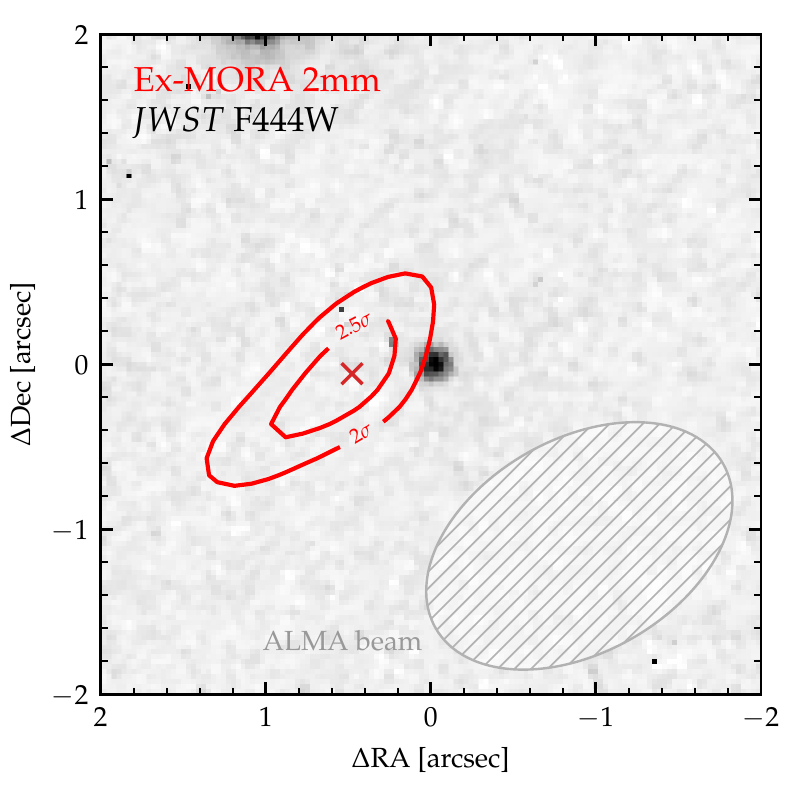}
\caption{Contours of the marginal 2 mm emission from Ex-MORA overlaid on the \JWST/F444W image. The location of the
$2.9\sigma$ peak is marked with an X, offset $\sim 0.5''$ from the NIRCam source but consistent within the positional uncertainty given the large beam size.
The Ex-MORA flux, if real, suggests a high IR luminosity of $\log L_{\rm IR}/L_\odot \sim 11.9^{+0.4}_{-0.3}$, comparable to local ULIRGs.
}\label{fig:fir_sed}
\end{figure}

Given that both \COS\ and \CEERS\ are very red, and our SED fits imply significant dust obscuration, they should be luminous in the FIR.
Both the COSMOS and EGS fields have coverage with \textit{Spitzer}/MIPS, \textit{Herschel}/PACS \& SPIRE, and JCMT/SCUBA-2, and we investigate these maps for any emission at the position of the NIRCam source. 
Specifically, we measure flux densities at the expected position of the source and find no significant emission in any of these FIR/submillimeter observations. 
This is consistent with expectations for a ULIRG at $z>7$ given the depth of these data.

However, \COS\ is covered in the Ex-MORA survey (A.~Long et al. \textit{in prep}), a blind 2 mm ALMA survey of the COSMOS field designed to identify high-redshift sub-mm galaxies \citep[an extension of the original MORA survey presented in][]{caseyMapping2021,zavalaEvolution2021}.\footnote{\CEERS\ is not covered by any archival NOEMA pointing in the EGS field.}
While Ex-MORA is still relatively shallow ($5\sigma \sim 1$ mJy), it is deeper than the available SCUBA-2 data and has a smaller beam size ($\theta_{\rm beam} \sim 1.5''$). 
At 2\,mm, such a dataset is optimal for identifying DSFGs at higher redshifts.
We find a marginal detection near the position of \COS, with a peak S/N of $2.9\sigma$ and flux density $S_{2\,{\rm mm}} = 0.19 \pm 0.07$ mJy offset by $\sim 0.5''$ from the NIRCam source. 
Figure~\ref{fig:fir_sed} shows contours of this marginal detection overlaid on the NIRCam/F444W image. 
While FIR positional offsets of $\sim 0.5$--$1.0$'' are not uncommon in high-redshift DSFGs \citep[e.g.~][]{biggsHighresolution2008, chapmanEvidence2004, hodgeEvidence2012, inamiALMA2022}, these are typically observed between the rest-frame UV and the FIR (rather than rest-frame optical and FIR) and in more extended sources.
However, we note that the 2 mm centroid position is uncertain to $\sim 0.5''$  \citep{condonErrors1997, ivisonSCUBA2007} due to the low S/N and $\sim 1.5''$ beam size, making it feasibly associated with the NIRCam source; as such, we do not consider the offset significant.

Though it requries follow-up verification, we examine the implications of this marginal detection for the FIR luminosity $L_{\rm IR}$. 
The only constraining power in the FIR SED comes from the marginal 2 mm Ex-MORA detection and the $850~\mu$m upper limit from SCUBA-2 ($3\sigma =2.9$ mJy).
We fit the FIR data to piecewise functions with a MIR power law and a FIR modified blackbody \citep[as in][]{caseyFarinfrared2012, drewNo2022}. 
We use a custom Markov Chain Monte Carlo (MCMC) routine \citep[based upon \texttt{MCIRSED};][]{drewNo2022} with flat priors on $L_{\rm IR}$, $T_{\rm dust}$, and $\beta$. 
In the absence of significant FIR constraints beyond our one 2 mm data point, we fix $\alpha_{\rm MIR} = 2.3$ and $\lambda_0$ = 200~$\mu$m \citep[following the recommendations in][]{drewNo2022}; we allow $T_{\rm dust}$ to vary from $26$ 
K ($\approx T_{\rm CMB}$ at $z=8.5$) to $90$ K and $\beta$ to vary from $1.5$ to $2.4$.
Due to the very negative $K$-correction in the sub-mm \citep*{caseyDusty2014}, the model SED is insensitive to the precise redshift in the 2 mm regime, so we adopt a fixed redshift of $z=8.5$. 
We account for the effects of heating by and decreasing contrast against the CMB following \citet{dacunhaEffect2013a}. 
We note that this MCMC fitting is not intended to constrain $T_{\rm dust}$ or $\beta$, but rather estimate $L_{\rm IR}$ marginalized over the uncertainty in the other SED parameters.

This fit yields an IR luminosity of $\log L_{\rm IR}/L_\odot \sim 11.9^{+0.4}_{-0.3}$. 
Based on the \citet{murphyCalibrating2011} calibration, this corresponds to an obscured SFR of $\sim 110^{+160}_{-60}\,{\rm M}_\odot\,{\rm yr}^{-1}$.
This implies a large fraction of obscured star-formation ($\sim 99$\%), as the upper limits in the rest-frame UV (uncorrected for dust) imply SFR$_{\rm UV} \lesssim 1~M_\odot~{\rm yr}^{-1}$. 
Moreover, we apply the methodology outlined in \citet[][see also \citealt{caseyPhysical2019} section 3.3]{scovilleISM2016} to estimate the dust mass in \COS. 
This method depends on the monochromatic flux at some wavelength (here 2 mm), the emissivity index (we assume $\beta=1.75$), and the \textit{mass-weighted} dust temperature, which is not the same as the luminosity-weighted temperature that can be derived in SED fitting. 
We adopt a temperature of 25~K, consistent with \citet{scovilleISM2016}. 
This yields a dust mass of $\log M_{\rm dust}/M_\odot = 8.2^{+0.5}_{-0.4}$, for $\beta \sim 1.5$--$2.4$.\footnote{We note that this method assumes the dust is optically thin, which is likely not the case at 2 mm (rest-frame $\sim 200\,\mu$m); however, in this case the dust mass would be underestimated.} 
This dust mass is consistent with the inferred attenuation $A_V \sim 1.5$--$2.5$, and more than an order of magnitude larger than the only known $z>8$ galaxy with a dust continuum detection \citep{tamuraDetection2019}. 
Deeper millimeter observations will be needed to verify this dust mass measurement and place constraints on the obscured SFR in this object.

\section{Discussion}\label{sec:discussion}

\textit{JWST} imaging is already leading to the identification of many interesting, high-$z$ candidates, some of which have unexpectedly red rest-frame optical colors. 
\citet{labbePopulation2023} presented a sample of red, $z_{\rm phot}>8$ candidates from early CEERS imaging which exhibited ``double breaks,'' both the Lyman and Balmer break. 
They interpret the red $2$--$4\,\mu$m color as tracing the 4000~\AA\ break and derive very large masses of $M_\star \gtrsim 10^{10}\,M_\odot$, potentially in excess of limits from $\Lambda$CDM cosmology \citep{boylan-kolchinStress2022a, menciHighredshift2022}. 
Recent work has suggested that the stellar masses of $z>10$ candidates may be overestimated due to differences in the IMF \citep[e.g.][]{steinhardtTemplates2022} or contamination from strong emission lines, e.g.~from AGN \citep[as discussed in][]{labbePopulation2023, endsleyJWST2022}.
Along those lines, \citet{kocevskiHidden2023} presented NIRSpec spectroscopy confirming one of the candidates in the \citet{labbePopulation2023} work as a $z=5.6$ broad-line AGN, which reduces their estimated volume density of massive $z>8$ galaxies.
\citet{furtakJWST2022} identify an extremely red, compact object at $z\sim 7.7$, similar to those in \citet{labbePopulation2023}, in deep \textit{JWST}/NIRCam imaging as part of the UNCOVER survey \citep{bezansonJWST2022}. 
Aided by lensing magnification, they constrain the effective radius to $\lesssim 35$ pc, providing strong evidence for a low-luminosity quasar with strong emission lines driving the red colors, though they do not rule out the possibility of highly dust-obscured star-formation.

\begin{figure*}
    \centering
    \includegraphics[width=\linewidth]{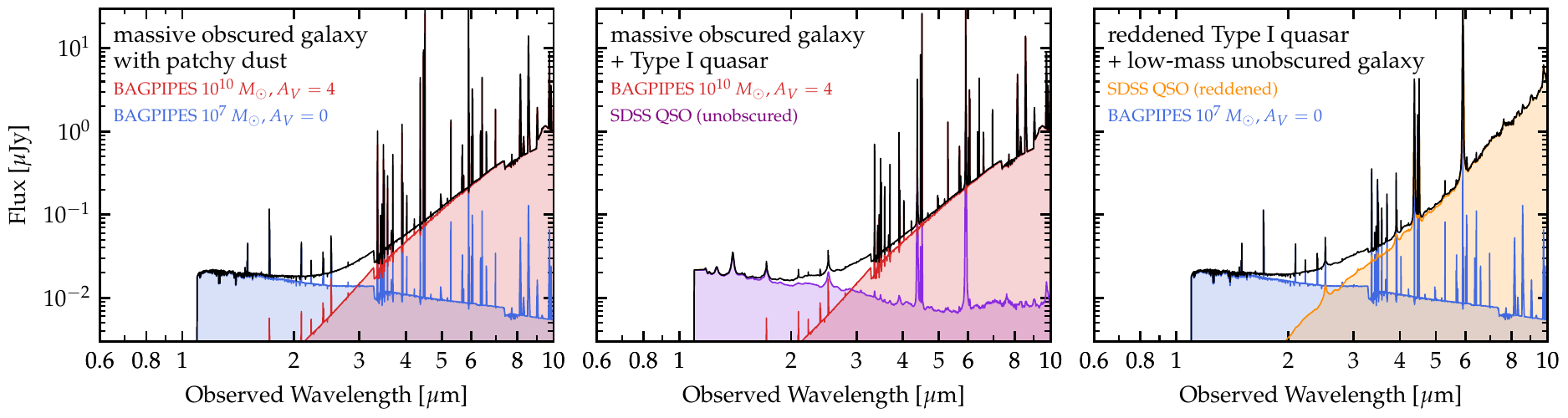}
    \caption{Illustration of the various galaxy+AGN possibilities associated with the composite SED shape of \CEERS, with a blue UV slope but red optical colors. \textbf{Left:} a galaxy-only composite model with a massive ($\log M_\star/M_\odot = 10$), obscured ($A_V=4$) galaxy combined with a low-mass ($\log M_\star/M_\odot = 7$) unobscured galaxy. Physically, this corresponds to a patchy distribution of dust with ``holes'' allowing some UV light to escape unattenuated. \textbf{Middle:} a galaxy+quasar model with a massive, obscured galaxy combined with a faint, unobscured Type I quasar. We adopt the SDSS QSO composite spectrum \citep{vandenberkComposite2001,glikmanNearInfrared2006}. \textbf{Right:} a galaxy+quasar model with a low-mass, unobscured galaxy combined with a bright but heavily dust-reddened Type I quasar.}\label{fig:schematic}
\end{figure*}

These results raise the question of whether these red, compact, high-$z$ objects are massive dust-obscured galaxies, reddened quasars, or both. 
Early massive black holes have indeed been invoked as a possible explanation for the high stellar masses inferred for the $z>9$ galaxy candidates being discovered by \textit{JWST} \citep{yuanRapidly2023, brummel-smithInferred2023}. 
Moreover, the objects presented in \citet{kocevskiHidden2023} and \citet{furtakJWST2022} show a red continuum at $\lambda_{\rm rest} > 3000$ \AA\ but a blue UV slope at shorter wavelengths, which has been interpreted as a composite galaxy+AGN signature.  
This SED shape is similar to \CEERS, and consistent with \COS\ given the shallower depth of the NIRCam data. 
Even with spectroscopic data, \citet{kocevskiHidden2023} do not come to a conclusion as to the origin of the two components. 
We show in Figure~\ref{fig:schematic} an illustration of the various galaxy/AGN possibilities consistent with this SED shape: in particular a galaxy-galaxy composite model (left), a blue quasar + red galaxy model (middle) and a blue galaxy + red quasar model (right). 
In the following subsections, we discuss the likelihood and implications of the two scenarios (dust-obscured galaxy vs.~quasar) for the candidates presented in this work.

\subsection{The high-z extreme of the DSFG population?}\label{sec:dsfg}

The $z>7$ regime is notoriously difficult for DSFG identification \citep{caseyAnalysis2018, caseyBrightest2018a} and therefore far-infrared spectroscopic follow-up has been limited to the brightest DSFGs \citep[with $L_{\rm IR} \sim 10^{13}\,L_\odot$][]{marroneGalaxy2018,endsleyALMA2022,fujimotoDusty2022}.
At the same time, far-IR follow-up of UV-luminous galaxies, which are far more numerous and more easily characterized than DSFGs, has found significant evidence for obscured star-formation at $z\sim 6$--$8$ \citep{schouwsSignificant2022, algeraALMA2023}.
These two samples represent two complementary approaches, from the rest-UV and the rest-FIR, to constraining the dust content of the early Universe.
However, constraints on the population in-between---obscured in the rest-UV, but not so bright as to be readily detectable in the IR---have been limited. 
The selection method presented in this paper, targeting extremely red objects in \JWST/NIRCam+MIRI imaging, represents an alternative approach. 

The two candidates presented in this paper appear to represent the high-redshift extreme of the population of moderately-obscured DSFGs at $z\sim 2$--$6$ \citep{chapmanRedshift2005, aretxagaAzTEC2011, smolcicMillimeter2012,mckinneyNearInfrared2023}.
This would place these sources among the progenitor population of the rare, extreme star-forming factories at $z\sim 5$--$7$ \citep[e.g.][]{zavalaDusty2018, marroneGalaxy2018, caseyPhysical2019, endsleyALMA2022}.
Indeed, the estimated volume density for these sources, $n\sim 2\times 10^{-5}\,$Mpc$^{-3}$, is comparable to the (albeit poorly constrained) volume density of $z\sim 7$ luminous infrared galaxies (LIRGS; $L_{\rm IR} > 10^{11}\,L_\odot$), for which integrating the IR luminosity function yields $n\sim 0.5$--$10\times 10^{-5}$ Mpc$^{-3}$ \citep{zavalaEvolution2021, fujimotoALMA2023}.

While neither candidate presented in this work is resolved in NIRCam/F444W imaging, such compact morphology is consistent with expectations for early galaxies. 
Indeed, to build up extreme stellar masses by $z\sim 8$ requires efficient funneling of gas into cold, dense clouds, and compact starbursts are often observed in bright sub-millimeter galaxies \citep{condonCompact1991, maSPT0346522016, jinDiagnosing2022}.
Theoretical work has suggested that massive galaxies at ultra-high-redshift may be able to form efficiently via feedback-free starbursts \citep{dekelEfficient2023}, which predicts compact, massive objects by $z\sim 8$--$10$.
Indeed, many $z>9$ galaxies being confirmed by \JWST\ are ultra-compact with effective radii $\lesssim 200$ pc \citep{tacchellaJADES2023,robertsonDiscovery2022a,roberts-borsaniShot2022,onoMorphologies2022}. 
We also note that the F444W photometry is likely dominated by the \oiii\ emission at $z\sim 7.5$--$8.5$, which may not be the best tracer of the morphology of the stellar continuum.

The composite blue+red SED of \CEERS\ could be due to a patchy distribution of dust in an overall very dust-obscured galaxy. 
In fact, many observed sub-millimeter galaxies (SMGs) at lower-redshift ($z\lesssim 5$; where we could efficiently probe the rest-UV continuum pre-\JWST) show blue UV slopes despite significant infrared excess \citep[e.g.][]{caseyAre2014}. 
Theoretical work has shown that a patchy dust geometry could allow a faint blue component to shine through ``holes'' in the ISM dust screen despite most of the stellar light being obscured \citep{poppingDissecting2017, narayananIRXv2018}. 
Indeed, we show in the left panel of Figure~\ref{fig:schematic} an illustration of the SED in this ``patchy dust'' scenario, in which a low-mass unobscured stellar population ($\log M_\star/M_\odot \sim 7$, $A_V = 0$) emits alongside a dust-obscured massive galaxy ($\log M_\star/M_\odot \sim 10$, $A_V \sim 4$). 
The fact that more and more $z>5$ galaxies are being identified with this unique SED shape is perhaps to be expected, as the sensitivity of \JWST\ allows us to observe the faint rest-UV emission from high-$z$ dust-obscured galaxies, which are expected to have complex star-dust geometry \citep{maDust2019}.

\subsubsection{Implications for dust production at $z>8$}

The interpretation of these candidates as high-$z$ dust-obscured galaxies would imply an early buildup of dust.
In particular, the marginal 2 mm flux from the Ex-MORA survey, if real and associated with \COS, suggests a high IR luminosity ($\log L_{\rm IR}/L_\odot = 11.9^{+0.3}_{-0.4}$) and high dust mass ($\log M_{\rm dust}/M_\odot = 8.2^{+0.5}_{-0.4}$). 
These estimates are highly uncertain and require direct sub-mm follow up for confirmation. 
However, if confirmed, they would have strong implications for the buildup of dust in the early Universe. 
The implied dust-to-stellar mass ratio of $M_{\rm dust}/M_\star \approx 0.03$ is significantly higher than low-redshift DSFGs \citep[e.g.][]{dunneHerschelATLAS2011}, but broadly consistent with the observed increase of this ratio at higher-redshift \citep{caluraDust2014, caluraDusttostellar2017}.

Furthermore, measurements of the dust mass at high redshift can provide stringent constraints on the relative contribution of different dust production mechanisms. 
There is abundant evidence that AGB stars, which are the dominant producers of dust later in cosmic time, are alone not able to produce large dust masses in $\lesssim 1$ Gyr \citep{valianteStellar2009, dwekOrigin2011, asanoDust2013};  
instead, dust production in SNe has been invoked to explain the large dust masses in $z>7$ galaxies. 
The recent detection of signatures of carbonaceous dust composition at $z=6.7$ \citep{witstokCarbonaceous2023} implies a large dust mass which requires either that significant star-formation occurred at $z>10$, or, more likely, that faster dust production channels dominate.

Even in the absence of FIR constraints, we can derive maximal dust masses based on the dust yield per AGB star or SN. 
In the dust-obscured galaxy interpretation, the stellar mass estimates for both sources in this paper are constrained thanks to MIRI imaging, making possible estimates of the number of AGB stars/SNe. 
Following the method outlined in \citet{michalowskiDust2015}, we estimate $N_{\rm AGB}$ and $N_{\rm SN}$ by integrating the IMF from $3$--$8\,M_\odot$ and $8$--$40\,M_\odot$, respectively. 
Here we assume a \citet{kroupaInitial2002a} IMF to be consistent with our \prospector-derived masses. 
We assume a theoretical maximum dust yield of $1.3\,M_\odot$ per SN\footnote{This assumes no dust destruction, likely an unphysical assumption. A more realistic yield of $\sim 0.1$--$0.15\,M_\odot$ per SN \citep{lesniewskaDust2019} would give a lower dust mass by a factor of $\sim 10$.} and $0.04\,M_\odot$ per AGB star \citep{michalowskiDust2015}. 
Based on the derived stellar masses of \COS\ and \CEERS\ (Table~\ref{tab:properties}), we derive maximum dust masses of $\sim 10^8\,M_\odot$ from SNe production and $\sim 10^7\,M_\odot$ from AGB stars. 

Other dust production mechanisms may therefore be needed to explain the high dust masses in these early galaxies, if confirmed. 
\citet{asanoDust2013} found that above a certain metallicity threshold ($\sim 0.3\,Z_\odot$), ISM grain growth can dominate over stellar dust production and can form $\sim 10^7\,M_\odot$ of dust in $\sim 100$ Myr.
Even more exotic, the maximal dust yields may be higher in unique cases such as dust produced in supershells \citep{martinez-gonzalezDust2021}, in the wake around Wolf-Rayet stars \citep{lauRevealing2021, lauNested2022}, in winds around an AGN accretion disk \citep{sarangiDust2019}, or in red-supergiant winds of high-mass Population III stars \citep{nozawaDust2014}. 
Moreover, a top-heavy IMF \citep[which has been suggested to be common in dust-obscured starbursts and high-$z$ star-formation in general, e.g.][]{zhangStellar2018} could increase the maximum dust mass by increasing the number of high-mass stars overall, especially for Population III stars. 
A higher early SNe rate may also drive accelerated grain growth in the ISM by the contribution additional seed metals at $z>10$. 
Taken altogether, these results highlight the importance of deep sub-mm follow-up of these objects and future samples of similarly-selected objects to constrain the dust masses and the physical processes responsible for the buildup of dust in the early universe.

\subsubsection{Implications for stellar mass assembly at $z\sim 8$}

The apparent ubiquity of massive ($M_\star > 10^{10}\,M_\odot$) galaxies identified at $z>8$ with \textit{JWST} has produced tensions with $\Lambda$CDM \citep{boylan-kolchinStress2022a, menciHighredshift2022}.
In the dust-obscured galaxy interpretation, the observed photometry implies stellar masses of $\sim 10^{10}\,M_\odot$ for \COS\ and \CEERS, even after correcting for the contribution from extremely bright emission lines.
Compared to the implied halo rarity, this suggests a high efficiency of converting baryons into stars, $\epsilon\sim 25\%$. 
This is consistent with results of \citet{inayoshiAssembly2020}, who derive constraints on the star-formation efficiency based on the $z>10$ candidates identified in early \textit{JWST} imaging \citep{harikaneSearch2022, naiduTwo2022, finkelsteinCEERS2022b, donnanEvolution2023a, harikaneComprehensive2023a}.
\citeauthor{inayoshiAssembly2020} note that, alternatively, such high stellar masses could be explained by a low baryon conversion efficiency in a metal-free stellar population with a top-heavy IMF. 
While this explanation may prove true for the UV-luminous population, the implied dust obscuration in the candidates presented here suggests a relatively metal-rich stellar population several generations beyond Population III. 
Therefore, if the stellar masses of these sources prove robust to constraints on any significant AGN contribution, they may suggest highly efficient stellar mass buildup in the early Universe. 

\begin{figure}
\centering
\includegraphics[width=\linewidth]{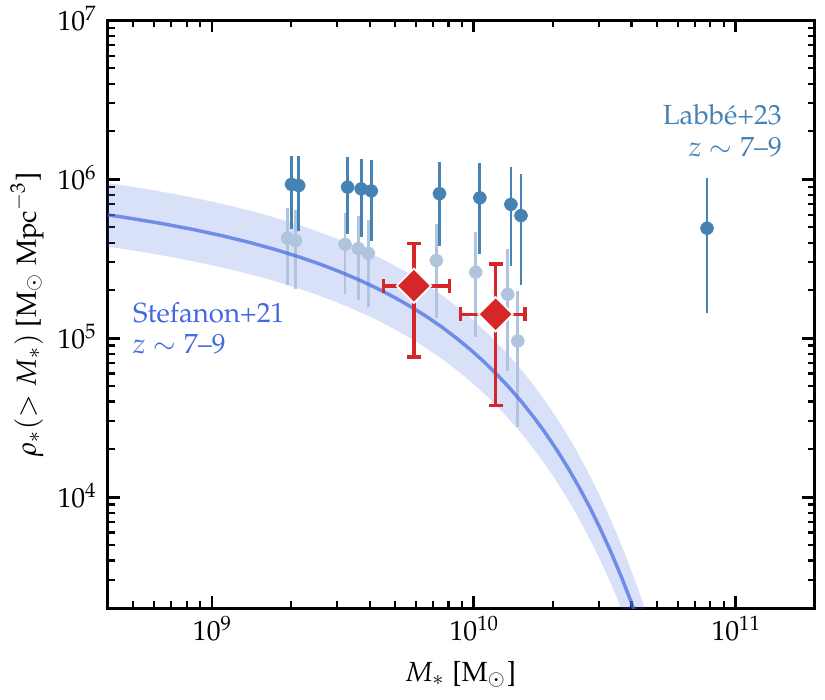}	
\caption{Cumulative stellar mass density $\rho_\star$ vs.~$M_\star$. The red points indicate the cumulative stellar mass density inferred from the detection of \COS\ and \CEERS.
We additionally plot the result for $z\sim 8$ from \citet{labbePopulation2023} and integrated Schechter function fits to UV-luminous LBGs from \citet{stefanonGalaxy2021a}.
In tabulating these results we adopt the same redshift bin $z\sim 7$--$9$ (see main text) and remove the $z=5.6$ AGN from the \citet{labbePopulation2023} sample. The light blue points show the \citet{labbePopulation2023} sample with the most massive candidate removed. The candidates presented in this work are consistent with the \citet{stefanonGalaxy2021a} $z\sim7$--$9$ SMF, though the uncertainties are large given the identification of just two sources. }\label{fig:rhostar}
\end{figure}

To explore the implications of these large stellar masses, we derive the cumulative stellar mass density $\rho_\star(>M_\star)$ vs. $M_\star$ based on the two candidates presented in this work.
We estimate the effective volume as the differential comoving volume integrated over a redshift bin from $z\sim 7$--$9$ and scaled to the total survey area of $20.6$ arcmin$^2$ (Table~\ref{tab:depths}).
Error bars include the cosmic variance uncertainty, which we compute from halo number counts in the DREaM simulation \citep{drakosDeep2022} as well as Poisson counting uncertainty. 
We note that the Poisson uncertainty dominates over cosmic variance ($\sim 70$--$90\%$ vs. $\sim 35\%$) given the detection of just two sources.

The resulting estimates are shown in Figure~\ref{fig:rhostar} alongside the equivalent measurements from \citet{labbePopulation2023}; we compute $\rho_\star$ in the same manner using their full $z\sim 7$--$9$ sample, after removing the $z=5.6$ AGN identified in \citet{kocevskiHidden2023}. 
The high stellar mass density from \citet{labbePopulation2023} is driven by the one $\log M_\star/M_\odot\sim 10.9$ candidate at $z\sim 7.5$; we show in light blue the $\rho_\star$ estimates with this candidate removed, just to highlight its impact on $\rho_\star$.  
The line and shaded region shows the stellar mass density corresponding to the Schechter fits stellar mass function from \citet{stefanonGalaxy2021a}, derived from samples of UV-luminous, IRAC-detected galaxies. 
To produce a curve comparable to our derived $\rho_\star$, we integrate the $\sim 7$, $8$, and $9$ SMFs and take a weighted mean. 
We weight each curve by the volume associated with redshift bins from $z\sim 7$--$7.5$, $7.5$--$8.5$, and $8.5$--$9$, respectively.

We find that the stellar mass density inferred from this work is consistent with the $z\sim 7$--$9$ stellar mass function inferred from the UV-luminous, IRAC-detected population. 
The large uncertainties prohibit a robust determination of the stellar mass density, and the relatively small area probed in this work prohibits constraints across a large dynamic range.
We note that the \citet{labbePopulation2023} sample is also fully consistent with the \citet{stefanonGalaxy2021a} SMF if the $\log M_\star/M_\odot \sim 10.9$ candidate were in fact lower stellar mass \citep[as suggested by][]{endsleyJWST2022}. 
Larger samples (e.g.~from the remaining COSMOS-Web and PRIMER imaging) are needed to constrain the contribution of obscured galaxies to the cosmic stellar mass density at this epoch.

\subsection{An early population of obscured AGN/reddened quasars?}\label{sec:agn}

Given the compact morphology and likely contribution of bright nebular lines, the observed emission in these candidates could be dominated by nuclear activity.
This could impact not just the EWs of optical emission lines but also potentially continuum emission. 
Based on size-mass and size-$z$ scaling relationships for known $z\sim 7$--$10$ star-forming galaxies \citep{onoEvolution2013, holwerdaSizes2015}, we might expect $R_{\rm eff} \sim 800$ pc at $M_\star \sim 10^{10}\,M_\odot$, significantly larger than the upper limit inferred from the F444W imaging ($\sim 200$ pc). 
However, this is by no means conclusive evidence for an AGN, as the dispersion in the size-mass relation is large at this epoch ($\sim 200$--$300$ pc), and (as discussed in \S\ref{sec:dsfg}) compact star-formation appears common at $z\sim 8$ \citep[e.g.][]{onoMorphologies2022}.

Given the lack of spectroscopy or rest-frame MIR coverage, the present data does not allow a robust constraint on any AGN contribution to the photometry. 
Here we discuss the different scenarios in which an AGN would impact the observed SED. 
First, a heavily obscured (i.e.~Type II) AGN would emit strongly in the rest-frame MIR due to hot torus dust. 
However, this is likely not significant shortward of rest-frame $\sim 2\,\mu$m, which is unconstrained by the present depth and filter coverage.  
Indeed, our \prospector-$\beta$ model includes emission from the dusty torus, but yields highly unconstrained values of $f_{\rm AGN}$ and $\tau_{\rm AGN}$. 
At this epoch, constraining the rest-frame MIR SED will be difficult and require ultra-deep MIRI imaging in the reddest wavelengths, which are also the least sensitive. 
Alternatively, deep X-ray or radio imaging could indicate the presence of an obscured AGN. 
However, neither source is detected in existing VLA radio images at $1.4$, $3$, and $5$ GHz \citep{schinnererVLACOSMOS2007, smolcicVLACOSMOS2017, ivisonAEGIS202007, willnerMidInfrared2006} or X-ray imaging from \textit{Chandra} \citep{lairdAEGISX2009, nandraAEGISX2015, civanoChandra2016, marchesiChandra2016}.

A second scenario involves strong emission lines from an AGN contaminating the broadband photometry \cite[e.g.][]{furtakJWST2022, endsleyJWST2022}. 
While we already account for the potential for strong \oiii\ and H$\alpha$ emission (albeit from star-forming \hii\ regions with slightly different emission line ratios) in our fits, strong emission lines from an AGN could also contribute significantly. 
To check this, we run \prospector\ including an empirical, scalable template for emission lines from the AGN narrow-line region \citep[NLR; based on data from][]{richardsonInterpreting2014}.
We first force the redshift to $z<7$ to examine the likelihood for lower-$z$ AGN interlopers \citep[i.e.][]{kocevskiHidden2023}; the resulting fits favor $z\sim 5$ but require unphysically high emission line EWs to match the observed photometry (\oiii+H$\beta$ EW $\sim 22000$ \AA) and even still achieve poor fits ($\chi_\nu^2 \sim 1.8$).

We then adopt a redshift prior based on the \prospector\ photometric redshifts reported in Table~\ref{tab:properties}.
We find consistent photometric redshifts and slightly ($\sim 0.2$ dex) lower stellar masses than our galaxy-only fits.
This difference in stellar mass comes largely from the fact that the current \prospector\ AGN NLR models do not include the associated nebular continuum emission, whereas the galaxy-only SED fits do. 
Regardless, as the MIRI 7.7\,$\mu$m flux is unaffected by the inclusion of strong AGN emission lines in the model, we consider this a robust constraint on the underlying continuum emission. 
Stellar mass estimates are therefore robust to this effect; however, emission line contribution from AGN would impact the current SFRs derived from SED fitting.
These AGN emission lines could be coming from an obscured Type II AGN or from an unobscured Type I AGN. 
In the latter case (i.e.~the picture presented in the middle panel of Figure~\ref{fig:schematic}), these sources may be in a transition stage between dust-enshrouded starbursts and unobscured luminous quasars \citep[e.g.][]{fuCircumgalactic2017,fujimotoDusty2022}.

Finally, given the compact nature of these sources and the relative expected rarity of $M_\star \sim 10^{10}~M_\odot$ systems, one might suspect that their reddened continuum emission is in fact not stellar in origin, but rather dominated by a highly reddened quasar. 
This interpretation (shown in the rightmost panel in Figure~\ref{fig:schematic}, and explored further in Barro et al.~\textit{in prep.}) would significantly impact the derived physical properties of the galaxy. 
The continuum in this case would be dominated by thermal emission from a dust-obscured accretion disk, with little to no contribution from a host galaxy, which would be completely sub-dominant. 
Unfortunately, constraining the relative contribution from stellar vs.~quasar continuum will require rest-frame mid-infrared diagnostics or deep radio/X-ray data, beyond what is feasible with current facilities.

Identification of such luminous reddened quasars in extremely low-mass galaxies would be unexpected: for example, the rightmost SED in Figure~\ref{fig:schematic} would have an implied black hole mass of $M_{\rm BH} \sim 10^7\,M_\odot$ (based on the $L_{\rm bol}$-$M_{\rm BH}$ relation) with a comparable host galaxy stellar mass of $M_\star \sim 10^7\,M_\odot$, the ratio of which is well outside of expectation \citep{mcconnellRevisiting2013}, even at high-$z$ \citep[e.g.][]{izumiSubaru2021b}.
The volume density of reddened Type I quasars is highly unconstrained. 
On the one hand, UV-luminous quasars are known to be very rare, $\sim 1000$ times rarer than these sources as inferred by integrating the \citet{matsuokaSubaru2018a} $z=6$ quasar luminosity function down to $M_{\rm UV} \sim -18$. 
At the same time, recent \JWST/NIRspec results have revealed an abundant population of broad-line AGN in $z\sim 7$ UV-faint galaxies \citep{harikaneJWST2023}. 
However, under the red QSO interpretation, the objects presented in this work are distinct from the populations of UV-luminous quasars (which are by definition unobscured by dust) or broad-line AGN (not all of which would be expected to dominate over the stellar continuum). 
The volume density of $z\gtrsim 7$ reddened quasars is likely somewhere in between the rare, UV-bright QSOs and the abundant broad-line AGN.

In summary, while these objects may host AGN, their measured stellar masses are robust to contribution from strong emission lines and hot dust torus emission. 
The major caveat is that we cannot rule out the possibility of continuum emission from dust-reddened Type I quasars.
However, we conclude that the dust-obscured galaxy interpretation is more likely based on the expected number densities of these classes of objects.
In particular, we note that the typical SF depletion time for $z\gtrsim 4$ DSFGs \citep[$\sim 100$--$300$ Myr;][]{swinbankALMA2014, aravenaALMA2016, manningCharacterization2022a, williamsDiscovery2019} is significantly longer than the typical quasar lifetime of $\sim 1$--$10$ Myr \citep{marconiLocal2004,volonteriCase2015,eilersDetecting2020}.
Follow-up spectroscopy will nevertheless be needed to search for AGN signatures (i.e.~broadened Balmer lines or weak high-ionization lines such as N\,\textsc{v} or C\,\textsc{iii}]).

\section{Summary} 

In this paper, we presented a search for extremely red, dust-obscured, $z>7$ galaxies in three publicly-available Cycle 1 surveys.
By focusing on sources detected in \textit{JWST}/NIRCam+MIRI imaging, we construct a unique selection for massive, red galaxies at $z>7$. 
\begin{enumerate}
	\item We identify two candidates, \COS\ and \CEERS, which have extremely red colors ($m_{277}-m_{444}$ $\sim 2.5$) and robust photometric redshifts of $8.5^{+0.3}_{-0.4}$ and $7.6^{+0.1}_{-0.1}$. The photometry for both sources is likely impacted by strong emission lines, particularly \oiii+H$\beta$ in F444W and H$\alpha$+\nii\ in F560W. Both candidates are significantly more dust obscured ($A_V\sim 2$--$3$) than other known $z\sim 8$ galaxies. 
    \item We find that neither source is resolved in NIRCam/F444W, constraining the rest-frame size to $R_{\rm eff} \lesssim 200$ pc. 
    \item We infer stellar masses of $\sim 10^{10}\,M_\odot$, significantly higher than known dust-continuum-detected galaxies at $z>8$ and similar to some of the most massive $z>8$ galaxy candidates yet identified by \JWST. The inferred stellar mass density is consistent within the uncertainty with expectations from the UV-luminous population. 
    \item We identify a marginal, $2.9\sigma$ detection at 2 mm near the position of \COS\ as part of the Ex-MORA survey. We show that this flux, if real, suggests an IR luminosity $\sim 10^{12}\,L_\odot$, consistent with the constraints on attenuation suggested by the \JWST\ data. There are no sub-mm constraints for \CEERS.
\end{enumerate}

This work highlights the importance of long wavelength MIRI imaging for characterization of massive, dust-obscured galaxies at $z>7$. 
Given the remarkable sensitivity of MIRI, almost $1$ mag deeper than pre-flight expectations \citep{caseyCOSMOSWeb2022b}, it becomes possible to constrain the $z>7$ galaxy population at rest-frame $\sim 1\,\mu$m. 
Future efforts to explore the dust-obscured population at this epoch will benefit greatly from deep MIRI imaging in multiple filters. 
While the completion of the full COSMOS-Web and PRIMER surveys will likely result in the detection of dozens more of these objects, spectroscopy and sub-mm follow-up will be necessary to determine the nature of the dust-obscured population in the epoch of reionization.

\section*{Acknowledgements}

Support for this work was provided by NASA through grant JWST-GO-01727 and HST-AR-15802 awarded by the Space Telescope Science Institute, which is operated by the Association of Universities for Research in Astronomy, Inc., under NASA contract NAS 5-26555.
H.B.A. acknowledges the support of the UT Austin Astronomy Department and thanks the UT Austin College of Natural Sciences for support through the Harrington Graduate Fellowship.
C.M.C. thanks the National Science Foundation for support through grants AST-1814034 and AST-2009577 as well as the University of Texas at Austin College of Natural Sciences for support; C.M.C. also acknowledges support from the Research Corporation for Science Advancement from a 2019 Cottrell Scholar Award sponsored by IF/THEN, an initiative of Lyda Hill Philanthropies.

Finally, H.B.A., C.M.C., and others at UT-Austin acknowledge that they work at an institution that sits on indigenous land. The Tonkawa lived in central Texas, and the Comanche and Apache moved through this area. We pay our respects to all the American Indian and Indigenous Peoples and communities who have been or have become a part of these lands and territories in Texas.

\facilities{\textit{HST} (ACS and WFC3), \textit{JWST} (NIRCam and MIRI), \textit{Spitzer} (IRAC and MIPS), \textit{Herschel} (PACS and SPIRE), SCUBA-2, VLA, ALMA. All the \JWST\ data used in this work can be found in MAST: \dataset[10.17909/4eps-dc89].}

\software{EAzY \citep{brammerEAZY2008}, \bagpipes\ \citep{carnallInferring2018a}, \prospector\ \citep{lejaDeriving2017a, johnsonStellar2021a}, \cigale\ \citep{boquienCIGALE2019},
\textsc{imfit} \citep{erwinIMFIT2015}, 
SourceExtractor \citep{bertinSExtractor1996},
SourceXtractorPlusPlus \citep{bertinSourceXtractor2020,kummelWorking2020},
\texttt{astropy} \citep{astropycollaborationAstropy2013}, 
\texttt{matplotlib} \citep{hunterMatplotlib2007}, \texttt{numpy} \citep{harrisArray2020}, \texttt{photutils} \citep{bradleyAstropy2022}, 
 \texttt{scipy} \citep{virtanenSciPy2020}, STScI JWST Calibration Pipeline \citep[\url{jwst-pipeline.readthedocs.io}][]{rigbyScience2022}.}

\bibliographystyle{aasjournal} 
\bibliography{z8_paper.bib}

\appendix

\vspace{-20pt}

\section{Additional candidate SEDs and photometry}\label{appendix:seds}
\counterwithin{figure}{section}
\setcounter{figure}{0}
\counterwithin{table}{section}
\setcounter{table}{0}

\begin{figure*}[h!]
    \centering    
    \includegraphics[width=.98\linewidth]{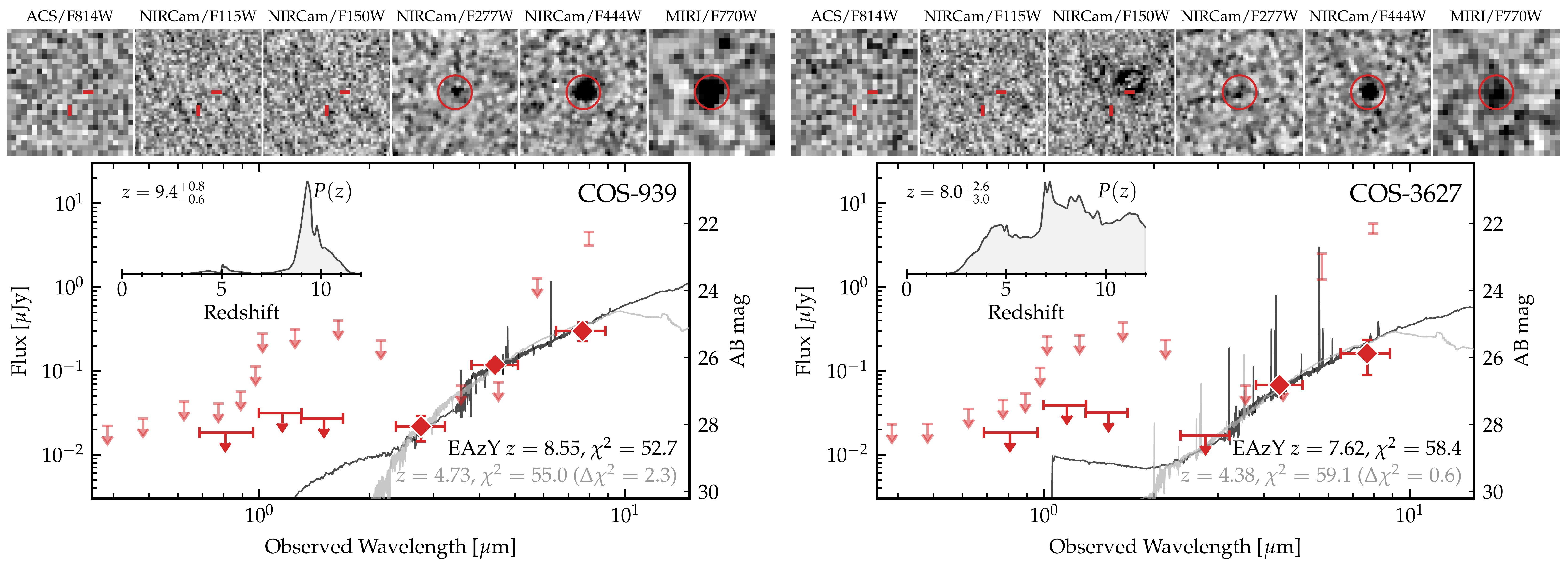}
    \caption{The optical-MIR SEDs of COS-939 and COS-3627, the two candidates that fall in our color-selection criteria but are fainter and thus have poorly constrained redshifts. As in Figures~\ref{fig:sed1} and \ref{fig:sed2}, we show cutouts in the ACS, NIRCam, and MIRI bands, and the measured photometry in red. We show the best-fit EAzY SED in black, and a forced $z<6$ model in grey.}%
    \label{fig:sed34}%
\end{figure*}

Table~\ref{tab:photometry} provides the photometric fluxes in the \HST/ACS, \JWST/NIRCam, and \JWST/MIRI bands for all objects in this paper. 
Figure~\ref{fig:sed34} shows the SEDs for COS-939 (left) and COS-3627 (right), the two candidates which fall into our color-color selection but have less robust SEDs due to redder $m_{444}-m_{770}$ colors (and general lower fluxes). 
We show the fiducial EAzY fit in black and a forced $z<6$ model in grey. 

The first candidate (COS-939) has a particularly red $m_{444}-m_{770}$ color ($\sim 1.5$) and drops out blueward of F277W yielding a redshift PDF that peaks at $z\sim 9$, but with a significant probability at $z\sim 5$ ($\sim 7\%$).
We note, however, that, COS-939 was initially identified as a potential close companion to \COS, located only $\sim 6''$ away ($\sim 30$ kpc at $z=8$). This potential association lends some credence to the possibility of a $z>7$ solution, but spectroscopic observations will be will be needed for confirmation.
The second candidate (COS-3627) suffers from a ``snowball'' artifact in the F150W image; this artificially elevated flux yields a blue $m_{150}-m_{277}$ color and drives the ultra high-$z$ solution as shown in Figure~\ref{fig:colorcolor}. 
When we mask this artifact and recompute photometry (as is shown in Fig.~\ref{fig:sed34}), we find an incredibly broad redshift PDF with nearly equal likelihood anywhere from $z\sim 4$ to $z\sim 12$---the source is too faint to achieve a robust constraint.

\begin{deluxetable*}{lccccccccccc}[b!]
\centering
\tablecaption{Photometry from \HST+\JWST. All fluxes are in nJy. \label{tab:photometry}}
\colnumbers
\tablewidth{700pt}
\tablehead{
	\colhead{\multirow{2}{*}{ID}} & \multicolumn{2}{c}{\HST/ACS} &\multicolumn{7}{c}{\JWST/NIRCam} & \multicolumn{2}{c}{\JWST/MIRI} \\[-0.7em] 
    & \colhead{F606W}& \colhead{F814W} & \colhead{F115W}& \colhead{F150W}& \colhead{F200W}& \colhead{F277W}& \colhead{F356W}& \colhead{F410M}& \colhead{F444W}& \colhead{F560W}& \colhead{F770W}\\[-2.1em]}
\startdata
\COS &\dots &  $(8 \pm 9)$ & $(22 \pm 25)$ & $(6 \pm 19)$ & \dots & $41 \pm 10$ & $150 \pm 33^{\dagger}$ & \dots & $307 \pm 11$ & \dots & $ 375 \pm 73$ \\
\CEERS & $(-1 \pm 5)$ & $(4\pm 5)$ & $11\pm 3$ & $(7\pm 4)$ & $(8\pm 3)$ & $17\pm 2$ & $69\pm 2$  & $158\pm 6$ & $188\pm 5$ & $451\pm 30$ & $338\pm 30$ \\
COS-939 & \dots & $(-6\pm 9)$ & $(-28\pm 16)$ & $(-2\pm 13)$ & \dots & $22\pm 7$ & $(-91\pm 33)^\dagger$ & \dots & $118\pm 8$ & \dots & $301\pm 73$ \\
COS-3627 & \dots & $(-3\pm 9)$ & $(16\pm 18)$ & $(0\pm 16)$ & \dots & $(14\pm 9)$ & $(15\pm 33)^\dagger$ & \dots & $68\pm 11$ & \dots & $162\pm 73$
\enddata
\tablenotetext{\dagger}{Reported flux is \textit{Spitzer}/IRAC [3.6].}
\end{deluxetable*}

\newpage
\allauthors

\end{document}